\documentclass[aps,prd,twocolumn,nofootinbib]{revtex4}

\usepackage{booktabs,makecell,multirow}
\usepackage{ulem}

\usepackage{graphicx,epsfig}
\usepackage{subfigure}

\usepackage[colorlinks=true, linkcolor=blue, citecolor=blue, urlcolor=blue]{hyperref}

\begin{document}

\title{A novel determination of non-perturbative contributions to Bjorken sum rule}

\author{Qing Yu}
\email{yuq@cqu.edu.cn}
\author{Xing-Gang Wu}
\email{wuxg@cqu.edu.cn}
\author{Hua Zhou}
\email{zhouhua@cqu.edu.cn}
\author{Xu-Dong Huang}
\email{hxud@cqu.eud.cn}

\affiliation{Department of Physics, Chongqing University, Chongqing 401331, People's Republic of China}
\address{Chongqing Key Laboratory for Strongly Coupled Physics, Chongqing 401331, P.R. China}

\date{\today}

\begin{abstract}

Based on the operator product expansion, the perturbative and nonperturbative contributions to the polarized Bjorken sum rule (BSR) can be separated conveniently, and the nonperturbative one can be fitted via a proper comparison with the experimental data. In the paper, we first give a detailed study on the pQCD corrections to the leading-twist part of BSR. Basing on the accurate pQCD prediction of BSR, we then give a novel fit of the non-perturbative high-twist contributions by comparing with JLab data. Previous pQCD corrections to the leading-twist part derived under conventional scale-setting approach still show strong renormalization scale dependence. The principle of maximum conformality (PMC) provides a systematic and strict way to eliminate conventional renormalization scale-setting ambiguity by determining the accurate $\alpha_s$-running behavior of the process with the help of renormalization group equation. Our calculation confirms the PMC prediction satisfies the standard renormalization group invariance, e.g. its fixed-order prediction does scheme-and-scale independent. In low $Q^2$-region, the effective momentum of the process is small and in order to derive a reliable prediction, we adopt four low-energy $\alpha_s$ models to do the analysis, i.e. the model based on the analytic perturbative theory (APT), the Webber model (WEB), the massive pQCD model (MPT) and the model under continuum QCD theory (CON). Our predictions show that even though the high-twist terms are generally power suppressed in high $Q^2$-region, they shall have sizable contributions in low and intermediate $Q^2$ domain. Based on the more accurate scheme-and-scale independent pQCD prediction, our newly fitted results for the high-twist corrections at $Q^2=1\;{\rm GeV}^2$ are, $f_2^{p-n}|_{\rm APT}=-0.120\pm0.013$, $f_2^{p-n}|_{\rm WEB}=-0.081\pm0.013$, $f_2^{p-n}|_{\rm MPT}=-0.128\pm0.013$ and $f_2^{p-n}|_{\rm CON}=-0.139\pm0.013$; $\mu_6|_{\rm APT}=0.003\pm0.000$, $\mu_6|_{\rm WEB}=0.001\pm0.000$, $\mu_6|_{\rm MPT}=0.003\pm0.000$ and $\mu_6|_{\rm CON}=0.002\pm0.000$, respectively, where the errors are squared averages of those from the statistical and systematic errors from the measured data.

\end{abstract}

\maketitle

\section{Introduction}

The Bjorken Sum Rule (BSR)~\cite{Bjorken:1966jh, Bjorken:1969mm}, which describes the polarized spin structure of nucleon, has been measured via polarized deep inelastic scattering (DIS) by various experimental collaborations~\cite{Anthony:1996mw, Abe:1994cp, Abe:1995mt, Abe:1995dc, Abe:1995rn, Abe:1997qk, Abe:1997dp, Abe:1998wq, Anthony:1999py, Anthony:1999rm, Anthony:2000fn, Anthony:2002hy, Adams:1994zd, Adams:1994id, Adams:1995ufa, Adams:1997hc, Adams:1997tq, Ackerstaff:1997ws, Ackerstaff:1998ja, Airapetian:1998wi, Airapetian:2002rw, Airapetian:2006vy, Alexakhin:2006oza, Alekseev:2010hc, Adolph:2015saz, Adolph:2016myg, Wesselmann:2006mw, Slifer:2008xu, Prok:2014ltt}. Using the operator product expansion (OPE), the BSR of the spin structure function can be calculated by separating the perturbative contribution of the matrix elements of local product operators from its non-perturbative contributions~\cite{Bjorken:1966jh, Bjorken:1969mm}, e.g.
\begin{eqnarray}
\Gamma^{p-n}_1(Q^2)&=&\int^1_0 dx[g^p_1(x, Q^2)-g^n_1(x, Q^2)]\nonumber\\
&=&\frac{g_A}{6}\left[1-E_{\rm ns}(Q^2)\right] +\sum\limits_{i=2}^{\infty}\frac{\mu_{2i}^{p-n}(Q^2)}{(Q^2)^{i-1}},
\label{gammapn}
\end{eqnarray}
where $g_1^{p,n}(x, Q^2)$ is the spin-dependent proton or neutron structure function with Bjorken scaling variable $x$, and $g_A$ is the nucleon axial charge. The BSR relates the difference of the proton and the neutron structure functions $\Gamma^p_1$ and $\Gamma^n_1$, and only the flavor non-singlet quark operators appear in perturbative part, resulting as the perturbative non-singlet leading-twist contributions $E_{\rm ns}(Q^2)$. The non-perturbative contribution is generally power suppressed in comparison to the leading-twist terms, which has been written as a power series over $1/Q^{2}$. Contributions from the high-twist terms could be sizable in low and intermediate $Q^2$ regions, and then the BSR provides a good platform for testing the perturbative and non-perturbative QCD contributions.

Analyses of $E_{\rm ns}(Q^2)$ under the $\rm{\overline{MS}}$-scheme have been given in the literature, such as Refs.\cite{Deur:2004ti, Deur:2008ej, Chen:2005tda, Deur:2014vea, Blumlein:2016xcy}. Additional treatment on extending the pQCD prediction to low $Q^2$-region has been done by using low-energy models for the strong coupling constant ($\alpha_s$) such as the analytic perturbation theory (APT), the ``massive analytic pQCD theory" (MPT), the $2\delta$- or $3\delta$-analytic QCD variants~\cite{Ayala:2018ulm, Ayala:2020scz, Pasechnik:2008th, Pasechnik:2009yc, Khandramai:2011zd, Khandramai:2013haz}. In all those treatments, there are large renormalization scale ($\mu_r$) dependence for the perturbative part due to the using of ``guessed" $\mu_r$; that is, in those analyses, the central (``optimal") value of $E_{\rm ns}(Q^2)$ is usually derived by setting $\mu_r = Q$, and then by varying it within an arbitrary range such as $[Q/2, 2Q]$ to estimate its uncertainty. Such guessing choice breaks the renormalization group invariance~\cite{Brodsky:2012ms, Wu:2014iba} and leads to conventional renormalization scale-and-scheme ambiguities due to the mismatching of the perturbative coefficients and the $\alpha_s$ at each order. In the literature, the principle of maximum conformality (PMC)~\cite{Brodsky:2011ta, Mojaza:2012mf, Brodsky:2012rj, Brodsky:2013vpa} has been suggested to eliminate such renormalization scale-and-scheme ambiguities. It is well known that the $\alpha_s$-running behavior is governed by the renormalization group equation (RGE). The existence of the $\{\beta_i\}$-terms emerged in the perturbative series is thus helpful for fixing exact $\alpha_s$-value of the pQCD approximant of a physical observable. And instead of choosing an optimal $\mu_r$, the PMC fixes the correct magnitude of $\alpha_s$ by using RGE, whose argument is called as the PMC scale, which is independent to any choice of $\mu_r$. The PMC prediction is scale-and-scheme independent, more detail and applications of the PMC can be found in the reviews~\cite{Wu:2013ei, Wu:2015rga, Wu:2019mky}.

To achieve a reliable prediction for the BSR high-twist contributions, it is important to have an accurate pQCD prediction on $E_{\rm ns}(Q^2)$. In the present paper, we shall first adopt the PMC single-scale approach~\cite{Shen:2017pdu} to deal with the perturbative part of the BSR, and then give a new determination of the non-perturbative high-twist contributions by comparing with the JLab data. The PMC singlet-scale approach follows the same idea of the original multi-scale approach~\cite{Brodsky:2011ta, Mojaza:2012mf, Brodsky:2012rj, Brodsky:2013vpa}, which determines an overall effective momentum flow of the process by using the RGE, whose magnitude corresponds to the weighted average of the multi-scales of the multi-scale approach at each order. It has also been demonstrated that the prediction under the PMC singlet-scale approach is scheme-and-scale independent up to any fixed order~\cite{Wu:2018cmb}. Though different from conventional scale ambiguity, there is residual scale dependence for fixed-order prediction due to unknown perturbative terms~\cite{Zheng:2013uja}. Such residual scale dependence can be greatly suppressed due to both $\alpha_s$-power suppression and exponential suppression. A detailed discussion on the residual scale dependence can be found in the recent review~\cite{Wu:2019mky}.

The remaining parts of the paper are organized as follows. In Sec.II, we present the calculation technology for the polarized Bjorken sum rule $\Gamma^{p-n}_1$. The PMC treatment of the pQCD contributions to the leading-twist part and the non-perturbative high-twist contributions shall be given. In Sec.III, we give the numerical results and discussions. Sec.V is reserved for a summary.

\section{Calculation technology}

In large $Q^2$-region, contributions from the leading-twist terms are dominant and those of the non-perturbative high-twist terms are generally power suppressed. In low and intermediate $Q^2$-region, contributions from the high-twist terms may have large contributions. In the following, we shall analyze the pQCD contributions to the leading-twist terms by using the PMC single-scale approach, and then give an estimation of the contributions from the non-perturbative high-twist terms. In low $Q^2$-region, the low-energy $\alpha_s$ models should be used; and for clarity, we shall adopt four low-energy $\alpha_s$ models to do our discussion.

\subsection{Perturbative series of the leading-twist terms}

The perturbative expansion over $\alpha_s$ for the hard part of the leading-twist terms $E_{\rm ns}(Q^2)$ has been calculated up to next-to-next-to-next-to leading order ($\rm{N^3LO}$), which can be written as
\begin{eqnarray}
E_{\rm ns}(Q^2,\mu_r)=\sum^4_{i=1}r_{i}(\mu_r)a^{i}(\mu_r),
\end{eqnarray}
where $a(\mu_r)=\alpha_s(\mu_r)/\pi$ and the perturbative coefficients $r_i$ are power series of the active flavor numbers $n_f$,
\begin{displaymath}
r_i =c_{i,0}+c_{i,1} n_f+\cdots +c_{i,n-1} n^{n-1}_f.
\end{displaymath}
The explicit expressions of the coefficients $c_{i,j}$ have been given in Refs.\cite{Baikov:2010je, Baikov:2012zm}. To apply the PMC, we need to use the general QCD degeneracy relations~\cite{Bi:2015wea} among different orders to make the transformation of the $n_f$-series to $\{\beta_i\}$-series, i.e. we need to rewrite $E_{\rm ns}(Q^2)$ in the following form,
\begin{eqnarray}
E_{\rm ns}(Q^2)&=&r_{1,0}a(\mu_r)+(r_{2,0}+\beta_0r_{2,1})a^2(\mu_r) \nonumber\\
&+&(r_{3,0}+\beta_1r_{2,1}+2\beta_{0}r_{3,1}+\beta_{0}^2r_{3,2})a^3(\mu_r) \nonumber\\
&+&(r_{4,0}+\beta_2r_{2,1}+2\beta_{1}r_{3,1}+\frac{5}{2}\beta_0\beta_1r_{3,2} \nonumber\\
&+&3\beta_0r_{4,1}+3\beta_0^2r_{4,2}+\beta_0^3r_{4,3})a^4(\mu_r) +\cdots,
\label{leadingtwist}
\end{eqnarray}
where the coefficients $r_{i,j}$ up to $\rm{N^3LO}$-order level are
\begin{eqnarray}
r_{1,0} &=& c_{1,0},  \\
r_{2,0} &=& c_{2,0} + {33\over2} c_{2,1},  \\
r_{2,1} &=& -6 c_{2,1},  \\
r_{3,0} &=&-{321\over8} c_{2,1}+ c_{3,0}+{33\over2} c_{3,1} + {1089\over 4} c_{3,2},  \\
r_{3,1} &=& {57\over4}c_{2,1}-3 c_{3,1}-99 c_{3,2},  \\
r_{3,2} &=& 36 c_{3,2},  \\
r_{4,0} &=& {11675\over 256}c_{2,1}-{321\over8}c_{3,1}-{10593\over8}c_{3,2}+c_{4,0}\nonumber\\
&&+{33\over2}c_{4,1}+{1089\over4}c_{4,2}+{35937\over8}c_{4,3}, \\
r_{4,1} &=& -{479\over16}c_{2,1}+{19\over2}c_{3,1}+{4113\over8}c_{3,2}-2 c_{4,1}\nonumber\\&&
-66c_{4,2}-{3267\over2}c_{4,3},\\
r_{4,2} &=& {325\over48}c_{2,1}-{285\over2}c_{3,2}+12c_{4,2}+594c_{4,3},\\
r_{4,3} &=& -216c_{4,3}.
\end{eqnarray}
Generally, the coefficients $r_{i, {j\neq0}}$ are functions of the logarithm ${\rm ln}(\mu_r^2/ Q^2)$. If setting $\mu_r=Q$, all those types of log-terms becomes zero, leading to a renormalon-free more convergent pQCD series; this explains why people usually choose $\mu_r=Q$ as the optimal scale for conventional scale-setting approach. Those coefficients can be reexpressed as
\begin{eqnarray}
r_{i,j} = \sum_{k=0}^{j} C_j^k \ln^k(\mu_r^2/Q^2) \hat{r}_{i-k,j-k},
\label{rij}
\end{eqnarray}
where the combination coefficients $C_j^k={j!}/{k!(j-k)!}$, and the coefficients $\hat{r}_{i,j}=r_{i,j}|_{\mu_r=Q}$. For convenience, we put the reduced coefficients $\hat{r}_{i,j}$ in the Appendix A.

The RGE, or the $\beta$-function, is defined as
\begin{equation} \label{RGEe}
\beta(a(\mu_r)) =-\sum_{i=0}^{\infty}\beta_{i}a^{i+2}(\mu_r).
\end{equation}
Following the decoupling theorem~\cite{Appelquist:1974tg}, the first two $\{\beta_{i\geq2}\}$-functions $\beta_{0}$ and $\beta_{1}$ are scheme-independent, and we have $\beta_0=\frac{1}{4}(11-\frac{2}{3}n_f)$ and $\beta_1=\frac{1}{4^2}(102-\frac{38}{3} n_f)$ for the ${\rm SU_{C}}(3)$-color group. The scheme dependent $\{\beta_{i\geq2}\}$-functions have been calculated up to five-loop level under the $\overline{\rm MS}$-scheme~\cite{Gross:1973id, Politzer:1973fx, Caswell:1974gg, Tarasov:1980au, Larin:1993tp, vanRitbergen:1997va, Chetyrkin:2004mf, Czakon:2004bu, Baikov:2016tgj}. A collection of all the known $\{\beta_{i}\}$-functions can be found in Ref.\cite{Wu:2013ei}.

In Eq.(\ref{leadingtwist}), the $\{\beta_i\}$-terms at each perturbative order govern the correct $\alpha_s$-running behavior, which inversely can be used to determine the effective magnitude of $\alpha_s$. Practically, by requiring all the RGE-involved non-conformal $\{\beta_i\}$-terms to be zero, one can achieve an overall effective $\alpha_s$ and hence the PMC scale $Q_\star$, and then the resultant pQCD series becomes the following scheme-independent conformal series:
\begin{equation}
E_{\rm ns}|_{\rm PMC}(Q^2) = \sum^{4}_{i\ge1} \hat{r}_{i,0} a^{i}(Q_\star),
\label{Enspmc}
\end{equation}
where the PMC scale $Q_{\star}$ can be fixed up to next-to-next-to-leading-log ($\rm NNLL$) accuracy by using the $\rm{N^3LO}$ perturbative series, e.g.
\begin{eqnarray}
\ln{\frac{Q_{\star}^2}{Q^2}}&=&  T _{0} +T_{1} {\alpha_{s}(Q)\over \pi}+T_{2} {\alpha_{s}^{2}(Q)\over \pi},
\label{Enspmcscale}
\end{eqnarray}
where
\begin{widetext}
\begin{eqnarray}
T_0 &=& -{\hat{r}_{2,1}\over \hat{r}_{1,0}}, \;\;
T_1 = {2(\hat{r}_{2,0}\hat{r}_{2,1}-\hat{r}_{1,0}\hat{r}_{3,1})\over  \hat{r}_{1,0}^2}  +{\hat{r}_{2,1}^2-\hat{r}_{1,0}\hat{r}_{3,2}\over \hat{r}_{1,0}^2}\beta_0,  \\
T_{2}&=&\frac{4 (\hat{r}_{1,0}\hat{r}_{2,0}\hat{r}_{3,1}-\hat{r}^2_{2,0}\hat{r}_{2,1})
+3(\hat{r}_{1,0}\hat{r}_{2,1}\hat{r}_{3,0}-\hat{r}^2_{1,0}\hat{r}_{4,1})}{ \hat{r}^3_{1,0}}+\frac{3(\hat{r}^2_{2,1}-\hat{r}_{1,0}\hat{r}_{3,2})}{2\hat{r}^2_{1,0}}
\beta_1\nonumber\\
&&-\frac{3\hat{r}_{2,0}\hat{r}^2_{2,1}- 4\hat{r}_{1,0}\hat{r}_{2,1} \hat{r}_{3,1}-2\hat{r}_{1,0}\hat{r}_{2,0}\hat{r}_{3,2}
+3\hat{r}^2_{1,0}\hat{r}_{4,2}}{\hat{r}^3_{1,0}}\beta_0 +\frac{2\hat{r}_{1,0}\hat{r}_{2,1}\hat{r}_{3,2} -\hat{r}^2_{1,0}\hat{r}_{4,3} -\hat{r}^3_{2,1}}{\hat{r}^3_{1,0}}\beta^2_0 .
\end{eqnarray}
\end{widetext}
Those equations show the PMC scale $Q_{\star}$ is exactly free of $\mu_r$, together with the $\mu_r$-independent conformal coefficients $\hat{r}_{i,0}$, the PMC prediction is exactly independent to any choice of $\mu_r$. Thus the conventional scale-setting ambiguity can be eliminated at any fixed-order by applying the PMC~\cite{Wu:2018cmb}. As a byproduct, due to the elimination of divergent renormalon terms in the resultant PMC perturbative series (\ref{Enspmc}), the pQCD convergence can be naturally improved. Those properties greatly improve the precision of the pQCD theory.

For a perturbative theory, it is important to have a reliable way to estimate the magnitude of the uncalculated higher-order terms. The scale-invariant and scheme-invariant PMC conformal series, which is also more convergent than the conventional series, is quite suitable for such purpose. A way of using the PMC series together with the Pad$\acute{e}$ approximation approach (PAA)~\cite{Basdevant:1972fe, Samuel:1992qg, Samuel:1995jc} has been suggested in Ref.\cite{Du:2018dma}. Some successful applications of this method can be found in Refs.\cite{Yu:2020tri, Huang:2020rtx, Yu:2019mce, Yu:2018hgw}. We shall adopt this method to estimate the magnitude of the unknown ${\cal O}(\alpha_s^5)$-terms of $E_{\rm ns}(Q^2)$, and in the following, we give a brief introduction of PAA.

The PAA offers a feasible conjecture that yields the $(n+1)_{\rm th}$-order coefficient by using a given $n_{\rm th}$-order perturbative series. For the purpose, people usually adopts a fractional function as the generating function. More explicitly, the $[N/M]$-type generating function of a pQCD approximant $\rho_n(Q)=\sum\limits_{i=1}^{n} \hat{r}_{i,0} a^{i}$ is defined as
\begin{eqnarray}
\rho^{[N/M]}_n(Q)  &=&  a \times \frac{b_0+b_1 a + \cdots + b_N a^N}{1 + c_1 a + \cdots + c_M a^M} \label{PAAseries0} \\
                   &=& \sum_{i=1}^{n} C_{i} a^{i} + C_{n+1}\; a^{n+1}+\cdots,   \label{PAAseries}
\end{eqnarray}
where $M\geq 1$ and $N+M+1=n$. The perturbative coefficients $C_i$ in Eq.(\ref{PAAseries}) can be expressed by the known coefficients $b_{i\in[0,N]}$ and $c_{j\in[1,M]}$. Inversely, if we have known the coefficients $C_i$'s up to $n_{\rm th}$-order level, one can determine the coefficients $b_{i\in[0,N]}$ and $c_{j\in[1,M]}$, and then achieve a prediction for the uncalculated $(n+1)_{\rm th}$-order coefficient $C_{n+1}$.

At the present, the leading-twist term $E_{\rm ns}|_{\rm PMC}$ has been known up to $\rm{N^3LO}$-level, and the four coefficients are known, $C_{i}=\hat{r}_{i,0}$ for $i\in[1, 4]$. Then the predicted ${\rm N^4LO}$-coefficient becomes
\begin{eqnarray}
\hat{r}_{5,0} =&& \frac{\hat{r}^4_{2,0} -3\hat{r}_{1,0} \hat{r}^2_{2,0} \hat{r}_{3,0} +\hat{r}^2_{1,0} \hat{r}^2_{3,0} +2\hat{r}^2_{1,0} \hat{r}_{2,0} \hat{r}_{4,0}}{\hat{r}^3_{1,0}},
\label{PAA03}
\end{eqnarray}
where the $[0/n-1]$-type PAA generating function has been implicitly adopted, which is the preferable type for the convergent PMC series~\cite{Du:2018dma}.

\subsection{Contributions from the non-perturbative high-twist terms}

The non-perturbative contributions to the BSR can be expanded in $1/Q^{2}$-power series as Eq.(\ref{gammapn}). The $\mathcal{O}(Q^{-2})$-term $\mu_4^{p-n}$ can be written as~\cite{Ji:1993sv, Shuryak:1981pi, Kawamura:1996gg}
\begin{eqnarray}
\mu_4^{p-n}=\frac{M^2}{9}(a^{p-n}_2 +4d^{p-n}_2 +4f^{p-n}_2),
\label{twist4}
\end{eqnarray}
where $M\approx0.94~\rm GeV$ is the nucleon mass. The leading-twist target mass correction $a^{p-n}_2$ can be calculated by using the leading-twist part of $g^{p-n}_1$, which is kinematically of high-twist~\cite{Blumlein:1998nv} and its magnitude at $Q^2=1~\rm{GeV^2}$ is $0.031\pm0.010$~\cite{Deur:2014vea}. The twist-3 matrix element $d^{p-n}_2$ is given by
\begin{eqnarray}
d^{p-n}_2=\int^1_0 dx x^2(2g_1^{p-n}+3g_2^{p-n}),
\end{eqnarray}
whose magnitude at $Q^2=1~\rm{GeV^2}$ is $0.008\pm0.0036$~\cite{Deur:2014vea}. The dynamical values of the twist-2 and twist-3 contributions can be measured by polarized lepton scattering off transversely and longitudinally polarized target. The twist-2 and twist-3 contributions are calculated by the $x^2$-weighted moment of the structure function in orders of $M^2/Q^2$, thus $a^{p-n}_2$ and $d^{p-n}_2$ change logarithmically, and we shall fix their values to be the above ones at $Q^2=1~\rm{GeV^2}$. Then the remaining undetermined term in $\mu_4^{p-n}$ is $f^{p-n}_2$. The twist-4 term $f^{p-n}_2$, which is related to the color electric and magnetic polarizabilities of nucleon, plays a pivotal role in phenomenological studies of the high-twist contributions. $f^{p-n}_2$ is sensitive to $Q^2$ and its $Q^2$-evolution satisfies~\cite{Shuryak:1981pi, Kawamura:1996gg}
\begin{eqnarray}
f^{p-n}_2(Q^2)=f^{p-n}_2(1)\left(\frac{a(Q)}{a(1)} \right)^{\gamma_0/8\beta_0},
\end{eqnarray}
where $\gamma_0/8\beta_0=32/81$ with $n_f=3$. The magnitude of $f^{p-n}_2(1)$ shall be fit by comparing with the data. Moreover, it has been argued that the $\mathcal{O}(Q^{-4})$-term $\mu_6^{p-n}$ may also have sizable contribution, so we take $\mu_6/Q^4$-term into consideration to have a better fit of the data.

\subsection{The strong coupling constant $\alpha_s$}

The $\alpha_s$-running behavior in perturbative region is governed by the RGE (\ref{RGEe}). Its solution can be written as an expansion over the inverse powers of the logarithm $L=\ln{\mu^{2}_{r} / \Lambda^2}$; and up to four-loop level, we have~\cite{Chetyrkin:1997sg}
\begin{eqnarray}
\alpha_s(\mu_r)&=&{\pi\over\beta_0L}\bigg\{1-{\beta_1\over\beta^2_0}{\ln L\over L}+{1\over\beta^2_0L^2} \left[{\beta^2_1\over\beta^2_0}(\ln^2L-\ln L  \right. \nonumber\\
&& \left. -1)+{\beta_2\over\beta_0} \right] +{1\over\beta^3_0L^3} \bigg[{\beta^3_1\over\beta^3_0}(-\ln^3L+{5\over2}\ln^2L\nonumber\\
&&+2\ln L-{1\over2})-3{\beta_1\beta_2\over\beta^2_0}\ln L+{\beta_3\over2\beta_0}\bigg] \bigg\},
\end{eqnarray}
where $\Lambda$ is the scheme-dependent asymptotic scale, which could be fixed by matching the measured value of $\alpha_s$ at a reference scale such as $M_Z$ or $m_\tau$ to its predicted value under a specific scheme.

In infrared region, when the scale is close to $\Lambda$ or even smaller, $\alpha_s$ becomes large whose magnitude cannot be well described by the RGE. To make the QCD prediction more reliable, we shall adopt four low-energy models for the $\alpha_s$ to do our calculation.

The first low-energy model is based on the analytical perturbation theory (APT)~\cite{Shirkov:1997wi, Shirkov:1997nx}, and we call it as the APT model. In APT model, its strong coupling constant $\alpha^{\rm{APT}}_s$ is described by applying the perturbation theory directly to the spectral function, which takes the following form,
\begin{eqnarray}
\alpha^{\rm{APT}}_s(\mu)&=&\frac{\pi}{\beta_0}\bigg(\frac{1}{\ln {\rm y}}+\frac{1}{1-{\rm y}}\bigg),
\label{asapt}
\end{eqnarray}
where $\mu$ is the energy scale, ${\rm y}={\mu^2}/{\Lambda^2}$ with
\begin{equation}
\Lambda^2=\mu^2 {\rm exp}[-\phi(\beta_0\alpha_s(\mu)/\pi)],
\label{Lambdaapt}
\end{equation}
where $\phi(z)$ satisfies $1/\phi(z)+1/(1-{\rm exp}[\phi(z)])=z$. Its freezing value is close to ${\alpha^{\rm{APT}}_s(10^{-10})/\pi}\approx0.43$.

The second low-energy model is an alteration of Eq.(\ref{asapt}), we call it as the WEB model~\cite{Webber:1998um}, which is suggested to suppress the nonperturbative power corrections of the APT model, and it takes the following form
\begin{eqnarray}
\alpha^{\rm WEB}_{s}(\mu)=\frac{\pi}{\beta_0}\bigg[\frac{1}{\ln {\rm y}}+\frac{{\rm y}+b}{(1-{\rm y})(1+b)}(\frac{1+c}{{\rm y}+c})^p\bigg],
\end{eqnarray}
where these phenomenological parameters $b=1/4$ and $p=c=4$. The obtained corresponding approximate freezing value $\sim \alpha^{\rm WEB}(10^{-10})/\pi\approx0.21$.

The third low-energy model is based on the ``massive analytic pQCD theory" (MPT)~\cite{Shirkov:1999hm, Shirkov:2012ux}, which takes the phenomenological glue-ball mass $m_{gl}=\sqrt{\xi}\Lambda$ as the infrared regulator, and we call it as the MPT model. It takes the following form
\begin{eqnarray}
\alpha^{\rm{MPT}}_s(\mu)&=&a_{cr}\bigg\{1+a_{cr}\frac{\beta_{0}}{\pi}\ln\left(1+ \frac{\mu^2}{m_{gl}^2}\right)+a_{cr}\frac{\beta_1}{\pi\beta_0}\times \nonumber\\
&&\ln\left[1+a_{cr}\frac{\beta_{0}}{\pi}\ln\left( 1+ \frac{\mu^2}{m_{gl}^2}\right)\right] +...\bigg\}^{-1},
\end{eqnarray}
whose freezing value at the origin satisfies $a_{cr}=\pi/(\beta_0 \ln\xi)$. Under the Landau gauge, we have $a_{cr}|_{\xi=10\pm2}=0.61\mp0.05$, which leads to the freezing point ${\alpha^{\rm{MPT}}_s(0)/\pi}=0.19^{-0.01}_{+0.02}$.

The fourth low-energy model is based on the continuum theory~\cite{Halzen:1992vd} and we call it as the CON model, where the exchanging gluons with effective dynamical mass $m_g$ is adopted and the non-perturbative dynamics of gluons is governed by the corresponding Schwinger-Dyson equation. It takes the following from
\begin{eqnarray}
\alpha^{\rm{CON}}_s(\mu)=\frac{\pi}{\beta_0\ln\left(\frac{4M^2_g+\mu^2}{\Lambda^2}\right)},
\end{eqnarray}
whose $M^2_g=m^2_g[\ln({\rm y}+4m^2_g/\Lambda^2)/\ln(4m^2_g/\Lambda^2)]^{-12/11}$ and $m_g=500\pm200$ MeV~\cite{Halzen:1992vd, cornwall:1982dy}, which leads to the freezing point ${\alpha^{\rm{CON}}_s(0)/\pi}=0.21^{-0.05}_{+0.19}$.

\section{Numerical results}

\begin{figure}[h]
\centering
\includegraphics[width=0.48\textwidth]{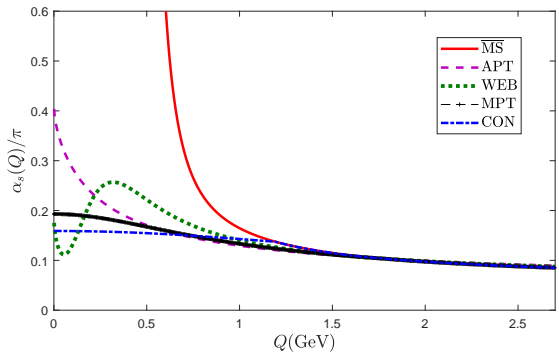}
\caption{Typical $\alpha_s$-running behavior in low-energy scales for four typical low-energy models, APT, WEB, MPT, and CON, respectively. The $\alpha_s$-running behavior derived from RGE under $\overline{\rm MS}$-scheme is given as a comparison.}
\label{coupling}
\end{figure}

To do the numerical analysis, we take the nucleon axial charge ratio $g_A=1.2724\pm0.0023$~\cite{PDG:2020}. The asymptotic QCD scale $\Lambda$ can be fixed by using the $\alpha_s$-value at the reference point such as $\alpha^{\overline{\rm MS}}_s(m_\tau)=0.325\pm0.016$~\cite{PDG:2020}, which gives $\Lambda_{\overline{\rm MS}}|_{n_f=3}=0.346^{+0.028}_{-0.029}$ GeV by using the four-loop RGE. Using the relation (\ref{Lambdaapt}), we obtain $\Lambda_{\rm APT}|_{n_f=3}=0.244^{+0.033}_{-0.031}$ GeV. In Fig.~\ref{coupling}, we present the typical running behaviors of $\alpha_s/\pi$ under four low-energy models, where the parameters are set to be $\xi=10$ for MPT and $m_g=700~\rm MeV$ for CON, respectively. The $\alpha_s$-running behavior derived from the RGE under $\overline{\rm MS}$-scheme is given as a comparison. Fig.~\ref{coupling} shows the importance of the using of low-energy models in the region of small energy-scale. Using the criteria suggested in Ref.\cite{Deur:2014qfa} for the analytic matching of $\alpha_s$ in perturbative and nonperturbative regimes, we obtain the transition scales ($Q_0$) for various low-energy models, which are $\sim 1.77$ GeV, $\sim 1.78$ GeV, $\sim 1.78$ GeV, $\sim 1.19$ GeV for APT, WEB,  MPT and CON models, respectively. As a subtle point, because the transition scales $Q_0$ for the cases of WEB and MPT are slightly bigger than $m_\tau$, and for self-consistency, we use the low-energy $\alpha^{\rm WEB/MPT}_s(m_\tau)=0.325\pm0.016$ to fix  $\Lambda$, which is $0.206\pm{0.022}$ GeV or $0.294^{+0.033}_{-0.032}$ GeV, respecitvely.

For later convenience, in the following discussions, we simply use $\alpha_s^{\overline{\rm MS}}$ to stand for the case of using $\overline{\rm MS}$-scheme $\alpha_s$ in all $Q^2$-region, $\alpha_s^{\rm APT}$ to stand for the case of using APT model in low-energy region ($Q<Q_0$, as mentioned above, $Q_0$ is different for different low-energy model) and $\overline{\rm MS}$-scheme $\alpha_s$ in large $Q^2$-region, $\alpha_s^{\rm WEB}$ to stand for the case of using WEB model in low-energy model and $\overline{\rm MS}$-scheme $\alpha_s$ in large $Q^2$-region, $\alpha_s^{\rm MPT}$ to stand for the case of using MPT model in low-energy model and $\overline{\rm MS}$-scheme $\alpha_s$ in large $Q^2$-region, and $\alpha_s^{\rm CON}$ to stand for the case of using CON model in low-energy model and $\overline{\rm MS}$-scheme $\alpha_s$ in large $Q^2$-region.

\subsection{Perturbative contributions to the leading-twist part of BSR up to $\rm{N^4LO}$ level}

\begin{figure*}[htb]
\centering
\subfigure[Leading-twist contributions using $\alpha^{\rm APT}_s$.]{
\includegraphics[width=0.45\textwidth]{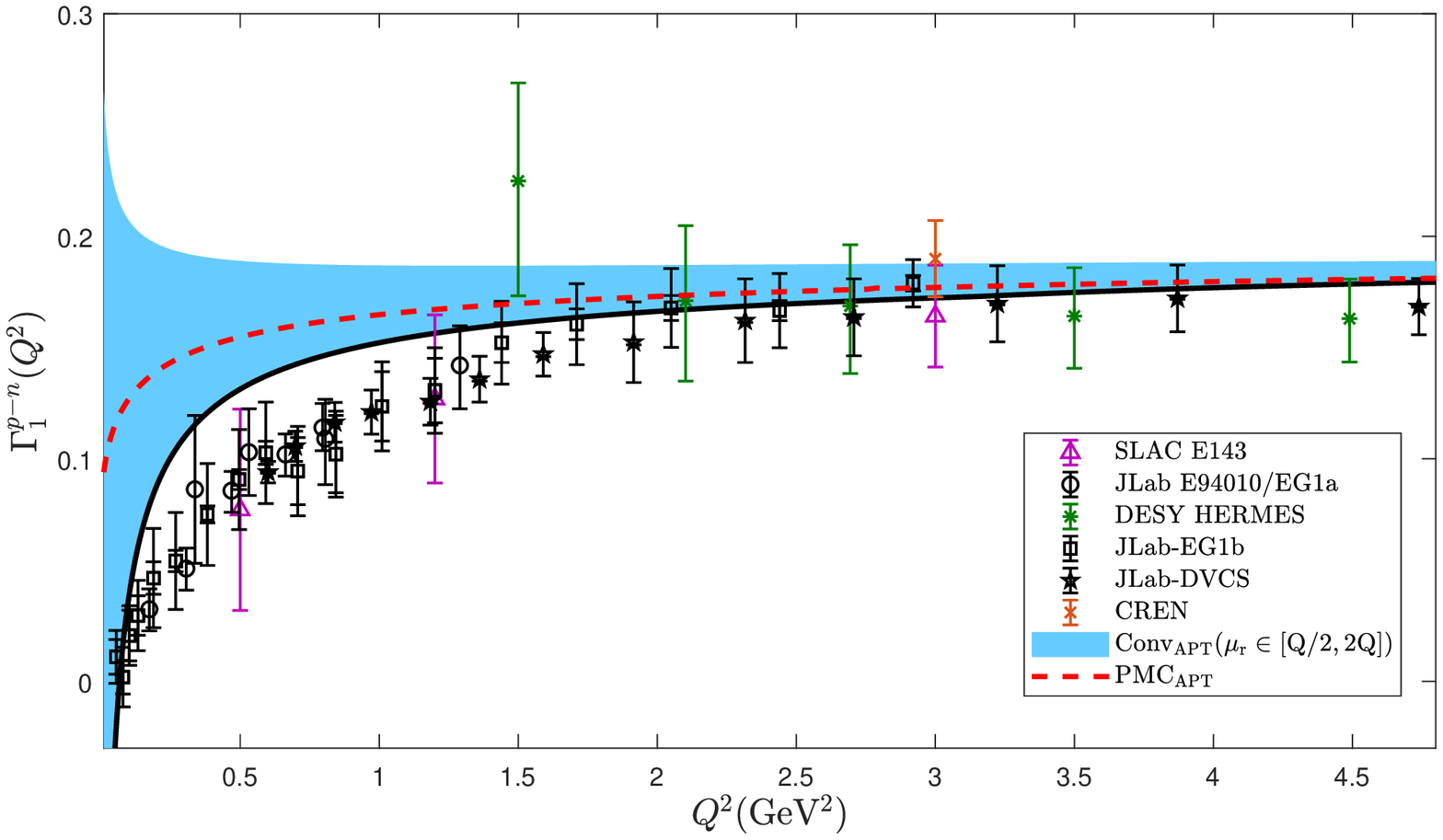}
}
\quad
\subfigure[Leading-twist contributions using $\alpha^{\rm WEB}_s$.]{
\includegraphics[width=0.45\textwidth]{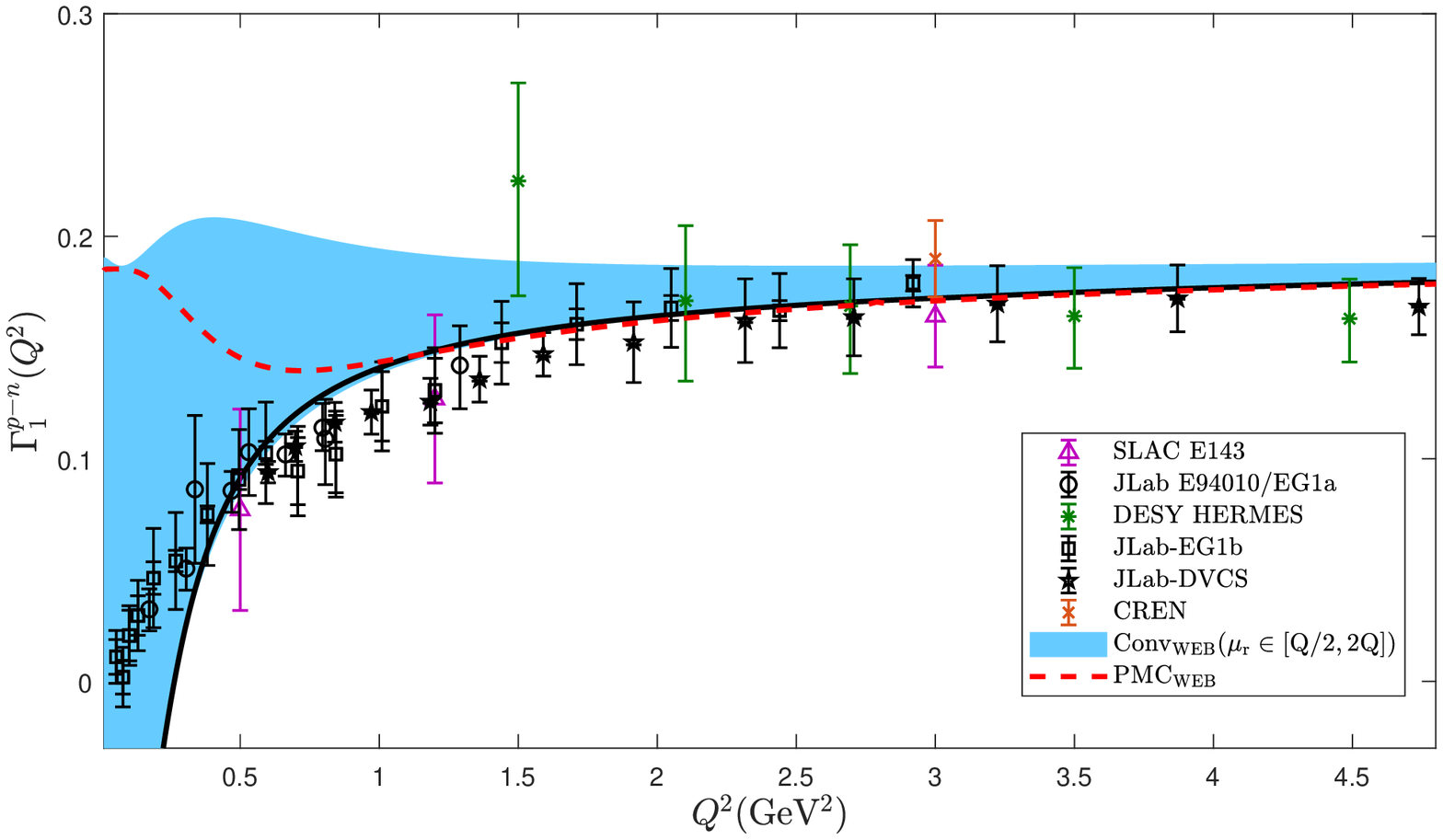}
}
\quad
\subfigure[Leading-twist contributions using $\alpha^{\rm MPT}_s$.]{
\includegraphics[width=0.45\textwidth]{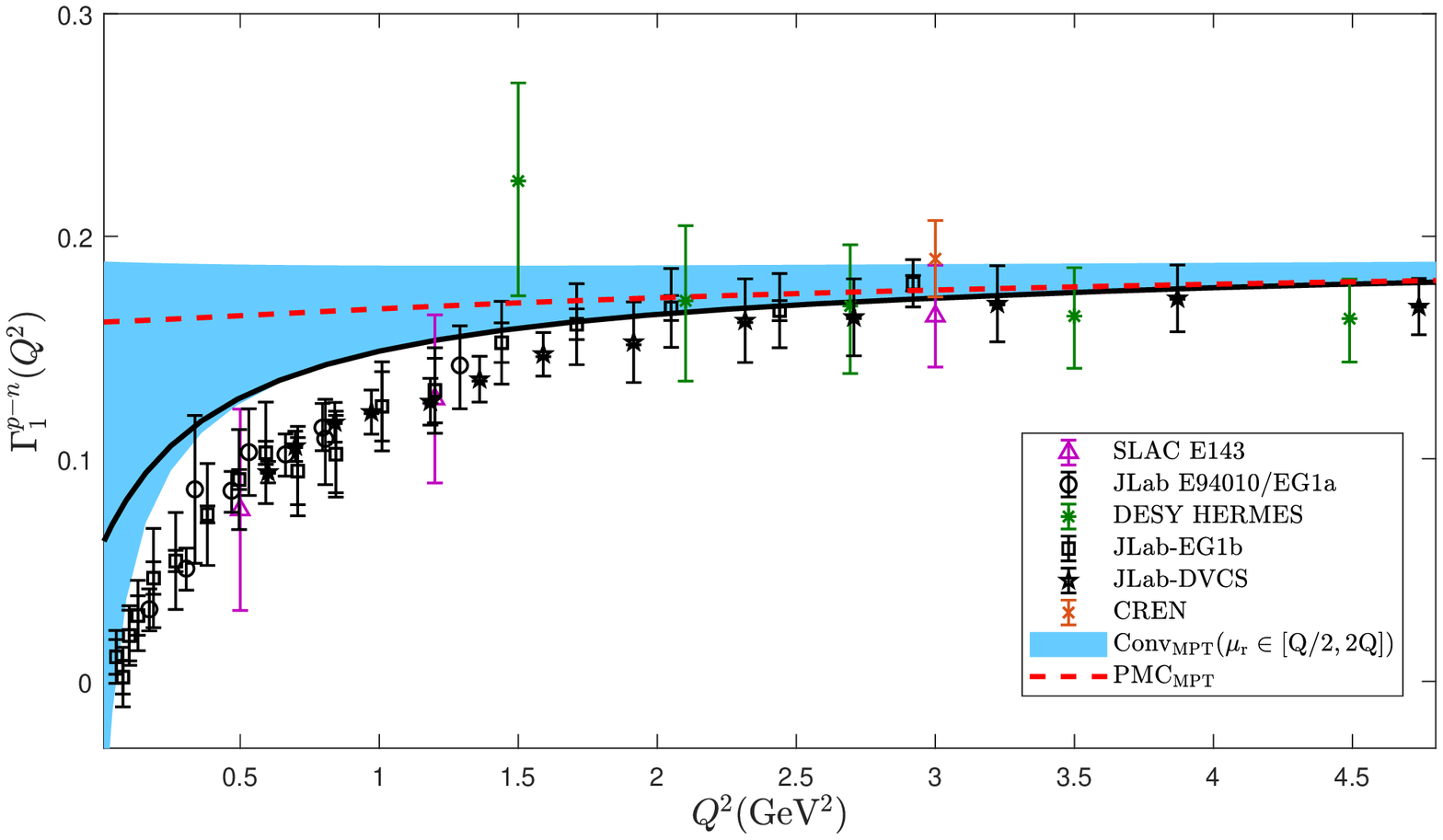}
}
\quad
\subfigure[Leading-twist contributions using $\alpha^{\rm CON}_s$.]{
\includegraphics[width=0.45\textwidth]{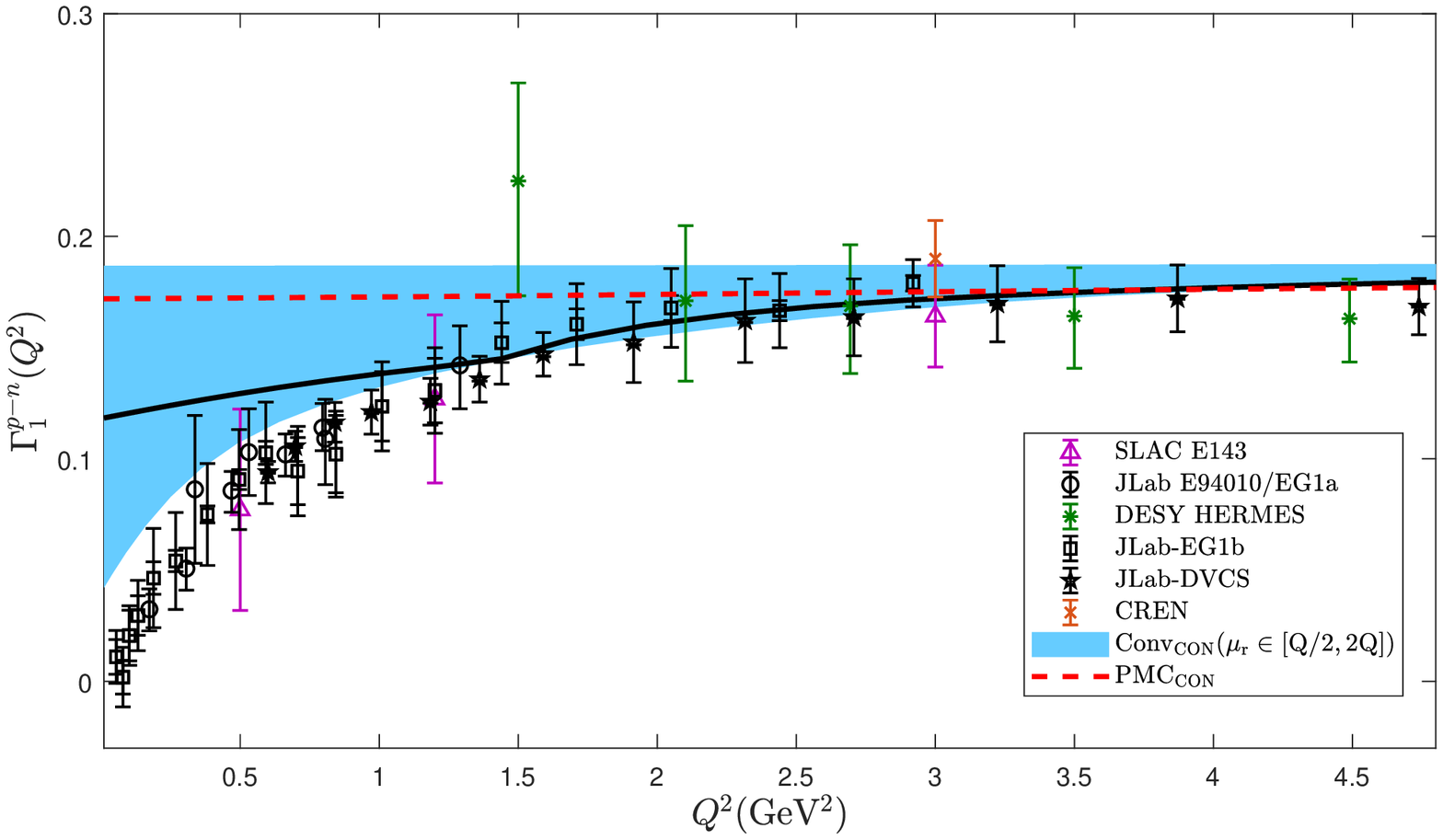}
}
\caption{Perturbative leading-twist contributions to the spin structure function $\Gamma^{p-n}_1(Q^2)$ up to $\rm{N^3LO}$ versus momentum Q, under four $\alpha_s$ models: (a) the APT model; (b) The WEB model; (c) the MPT model; and (d) the CON model. The solid line is for conventional scale setting approach with $\mu_r=Q$ and the shaded band shows its scale uncertainty by varying $\mu_r\in[Q/2,2Q]$. The dot-dashed line is the prediction $\Gamma^{p-n}_1(Q^2)$ up to $\rm{N^4LO}$ for PMC scale-setting approach, which is free of renormalization scale dependence.}
\label{leadingBSR}
\end{figure*}

The perturbative contributions to the leading-twist part $E_{\rm ns}(Q^2)$ has been known up to $\rm{N^3LO}$. Under conventional scale-setting approach, the pQCD series is scale dependent, and by setting $\mu_r=Q$, we obtain
\begin{eqnarray}
E_{\rm ns}(Q^2)|_{\rm Conv.}&=& a(Q) +3.58 a^2(Q) +20.22 a^3(Q)\nonumber\\
&&+175.70 a^4(Q).
\label{EnscoefficientsC}
\end{eqnarray}
On the other hand, the pQCD series becomes scale invariant by applying the PMC, and we obtain
\begin{eqnarray}
E_{\rm ns}(Q^2)|_{\rm PMC}&=& a(Q_\star) +1.15 a^2(Q_\star) +0.14 a^3(Q_\star) \nonumber\\
&&+0.76 a^4(Q_\star)
\label{EnscoefficientsS}
\end{eqnarray}
for any choice of renormalization scale, where $Q_\star$ is of perturbative nature, which can be determined up to NNLL accuracy
\begin{eqnarray}
\ln\frac{{Q_{*}}^2}{Q^2}&=&-1.08 -1.87 a(Q) -24.06 a^2(Q).
\end{eqnarray}

One may observe that the perturbative coefficients in PMC series (\ref{EnscoefficientsS}) are much smaller than those of conventional series (\ref{EnscoefficientsC}), especially for those of high-orders, which are due to the elimination of divergent renormalon terms as $n!\beta^n_0 a_s^n$. This indicates that a much more convergent perturbative series can be achieved by applying the PMC. At the same time, the PMC scale $Q_\star$ also shows a fast convergent at high $Q$-range, e.g. the relative absolute values of the LL, the NLL and the NNLL terms are 1: 0.064 : 0.030 for $Q=100$ GeV. Thus the residual scale dependence due to unknown even higher-order terms can be greatly suppressed.

Using the convergent PMC perturbative series, one can obtain a reliable prediction of unknown $\mathcal{O}(a^5)$-term by using the PAA, e.g. by using Eq.(\ref{PAA03}), we obtain
\begin{equation}
 E_{\rm ns}(Q^2)|^{\rm N^4LO}_{\rm PAA}=2.92 a^5(Q_\star).
\end{equation}

We present the predicted leading-twist part of the spin structure function $\Gamma^{p-n}_1(Q^2)$ under four low-energy models in Fig.~\ref{leadingBSR}, where the results under conventional and PMC scale-setting approaches are presented. The experimental data are from SLAC~\cite{Abe:1994cp, Abe:1995mt, Abe:1995dc, Abe:1995rn, Abe:1998wq}, DESY~\cite{Ackerstaff:1997ws, Ackerstaff:1998ja, Airapetian:1998wi, Airapetian:2002rw, Airapetian:2006vy}, CREN~\cite{Alexakhin:2006oza, Alekseev:2010hc, Adolph:2015saz} and JLab~\cite{Deur:2004ti, Deur:2008ej, Deur:2014vea}. The PMC predictions are independent to any choice of $\mu_r$, and the shaded band shows the conventional renormalization scale uncertainty by varying $\mu_r\in[Q/2, 2Q]$. Under conventional scale-setting approach, the spin structure function $\Gamma^{p-n}_1(Q^2)$ shows large scale dependence, especially in low-energy region. In low-energy region, the results by using the IR-fixed couplings are much more reliable. And since couplings behaves differently in low-energy region, the spin structure function $\Gamma^{p-n}_1(Q^2)$ behaves quite differently for $Q\to 0$. When the energy scale is large enough, such as $Q> 1.5-2.0$ GeV, the perturbative leading-twist terms could explain the experimental data well. Fig.~\ref{leadingBSR} also shows that in low-scale region, the leading-twist terms alone cannot explain the data and one must take the high-twist terms into consideration. By comparing with the data, this fact inversely provides us a good platform to achieve reliable predictions on the magnitudes of high-twist contributions.

\subsection{Analysis of high-twist contributions under various low-energy models}

\begin{table*}[htb]
\centering
\begin{tabular}{  c c  c  c c c c c c}
\hline
 & ~$\alpha_s$ models ~   &~   & ~~~$f_2^{p-n}(1)$~~~ &~~~$\mu_6$~~~ & ~~~$\chi^2/d.o.f$ \\
\hline
&~                     &$\mu_r=Q/2$          &~~$-0.176\pm{0.000}\pm{0.013}$~~ & ~~$0.004\pm{0.000}\pm{0.000}$~~ &149 \\
&$\rm{APT}|_{Conv}$    &$\mu_r=Q$            &$-0.088\pm{0.000}\pm{0.013}$  &$0.002\pm{0.000}\pm{0.000}$ &62 \\
&~                     &$\mu_r=2Q$           &$-0.107\pm{0.000}\pm{0.013}$ &$0.003\pm{0.000}\pm{0.000}$ &117 \\
&$\rm{APT|_{PMC}}$     &$\mu_r\in[Q/2,2Q]$   &$-0.120\pm{0.000}\pm{0.013}$ &$0.003\pm{0.000}\pm{0.000}$  &62 \\
\hline
&~                     &$\mu_r=Q/2$          &$-0.193\pm{0.000}\pm{0.013}$ &$0.005\pm{0.000}\pm{0.000}$ &193 \\
&$\rm{WEB}|_{Conv}$    &$\mu_r=Q$            &$-0.047\pm{0.000}\pm{0.013}$  &$0.001\pm{0.000}\pm{0.000}$ &168  \\
&~                     &$\mu_r=2Q$           &$-0.105\pm{0.000}\pm{0.013}$ &$0.004\pm{0.000}\pm{0.000}$ &160 \\
&$\rm{WEB|_{PMC}}$     &$\mu_r\in[Q/2,2Q]$   &$-0.081\pm{0.000}\pm{0.013}$ &$0.001\pm{0.000}\pm{0.000}$  & 45 \\
\hline
&~                     &$\mu_r=Q/2$          &$-0.173\pm{0.000}\pm{0.013}$ &$0.004\pm{0.000}\pm{0.000}$ &151 \\
&$\rm{MPT}|_{Conv}$    &$\mu_r=Q$            &$-0.080\pm{0.000}\pm{0.013}$ &$0.002\pm{0.000}\pm{0.000}$ &56\\
&~                     &$\mu_r=2Q$           &$-0.105\pm{0.000}\pm{0.013}$ &$0.003\pm{0.000}\pm{0.000}$ &126\\
&$\rm{MPT|_{PMC}}$     &$\mu_r\in[Q/2,2Q]$   &$-0.128\pm{0.000}\pm{0.013}$ &$0.003\pm{0.000}\pm{0.000}$ &50 \\
\hline
&~                     &$\mu_r=Q/2$         &$-0.175\pm{0.000}\pm{0.013}$ &$0.003\pm{0.000}\pm{0.000}$ &125 \\
&$\rm{CON}|_{Conv}$    &$\mu_r=Q$           &$-0.070\pm{0.000}\pm{0.013}$ &$0.001\pm{0.000}\pm{0.000}$ &60 \\
&~                     &$\mu_r=2Q$          &$-0.102\pm{0.000}\pm{0.013}$ &$0.002\pm{0.000}\pm{0.000}$ &138 \\
&$\rm{CON|_{PMC}}$     &$\mu_r\in[Q/2,2Q]$  &$-0.139\pm{0.001}\pm{0.013}$ &$0.002\pm{0.000}\pm{0.000}$ &49 \\
\hline
\end{tabular}
\caption{The fitted parameters $f^{p-n}_2(Q^2=1~\rm GeV^2)$ and $\mu_6$ and their corresponding quality of fit $\chi^2/d.o.f$ under four $\alpha_s$ models before and after applying the PMC, where the first and the second errors are caused by the statistical and systematic errors of the data~\cite{Deur:2008ej, Deur:2014vea}. The twist-6 coefficient $\mu_6$ is almost independent to the choices of statistical and systematic errors.}
\label{fitf2mu6}
\end{table*}

Following the discussions of Sec.II.B, we need to fit two parameters, $f^{p-n}_2(1~\rm GeV^2)$ and $\mu_6$, so as to determine the high-twist contributions. We adopt the most recent data listed in Refs.\cite{Deur:2008ej, Deur:2014vea} to do the fitting, whose momentum transfer lies in the range of ${0.054}~{\rm GeV^2} \leq Q^2\leq 4.739~{\rm GeV^2}$. We adopt the APT, WEB, MPT, and the CON couplings in doing the fitting. The quality of fit is measured by the parameter of $\chi^2/ d.o.f$, e.g.
\begin{equation}
\chi^2/{d.o.f} = {1\over {N-d}}\sum\limits^{N}_{j=1}\frac{(\Gamma^{p-n}_{1, {\rm the.}}(Q^2_j) -\Gamma^{p-n}_{1,{\rm exp.}}(Q^2_j))^2}{\sigma^2_{j,{\rm stat.}}},
\end{equation}
where the symbol ``$d.o.f$" (short notation of the degree of freedom) is equal to $N-d$ with $N=31$ being the number of data points and $d=2$ being the number of fitted parameters, ``the." stands for theoretical prediction, ``exp." stands for measured value, and ``$\sigma_{j,{\rm stat.}}$" is the statistical error at each point $Q_j$. Comparing theoretical prediction $\Gamma^{p-n}_{1, {\rm the.}}(Q^2_j)$ with the measured value $\Gamma^{p-n}_{1, {\rm exp.}}(Q^2_j)$ at all the data points $Q_{j\in[1,N]}$, we can derive the preferable $f^{p-n}_2$ and $\mu_6$ by requiring them to achieve the minimum value of $\chi^2/d.o.f$. To do the fitting, we also take into account the systematic error $\sigma_{j, {\rm sys.}}$ at each point $Q_j$, which has sizable contributions to the fitted values of $f^{p-n}_2$ and $\mu_6$. For convenience, we put the detailed calculation technology in Appendix B.

Our results for the two parameters $f^{p-n}_2(1~\rm GeV^2)$ and $\mu_6$ are presented in Table.~\ref{fitf2mu6}. The right-most column shows the smallest $\chi^2/d.o.f$ for the predictions before and after applying the PMC under four $\alpha_s$ models. The magnitudes of those two parameters are small, which agree with the usual consideration that at large $Q^2$-region, the high-twist terms are power suppressed and are negligible. However in low $Q^2$-region, they will have sizable contributions; especially $f^{p-n}_2(1~\rm GeV^2)$ is important for a reliable theoretical prediction on $\Gamma^{p-n}_{1, {\rm the.}}(Q^2)$ in low $Q^2$-region. Table.~\ref{fitf2mu6} shows that the fitted parameters under conventional scale-setting approach have strong scale dependence, whose quality of fit $\chi^2/d.o.f$ varies from tens to hundreds, and the optimal fit are achieved for the case of $\mu_r\sim Q$. This, together with a better pQCD convergence due to the elimination of divergent log-terms $\ln\mu_r^2/Q^2$, in some sense explain why $\mu_r=Q$ is usually taken as the preferable renormalization scale for conventional scale-setting approach. On the other hand, the fitted parameters for the PMC scale-setting approach is independent for any choice of renormalization scale, thus a more reliable and accurate prediction is achieved.

\begin{figure}[htb]
\centering
\includegraphics[width=0.48\textwidth]{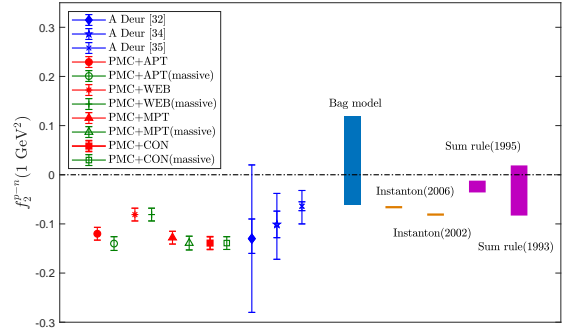}
\caption{The twist-4 coefficient $f_2^{p-n}(1{\rm~GeV^2})$ obtained from the PMC predictions under four $\alpha_s$ low-energy models, in which the predictions using JLab data~\cite{Deur:2004ti, Deur:2008ej, Deur:2014vea}, the QCD sum rule predictions~\cite{Stein:1995si, Balitsky:1989jb}, and the predictions using the model of the instanton-based QCD vacuum~\cite{Lee:2001ug, Sidorov:2006vu} and the Bag model prediction~\cite{Ji:1993sv} are also presented. }
\label{f2pn}
\end{figure}

At present,  the twist-4 coefficient $f^{p-n}_2(1~\rm GeV^2)$ has been calculated under various approaches, such as Refs.~\cite{Stein:1995si, Balitsky:1989jb, Lee:2001ug, Sidorov:2006vu, Ji:1993sv, Deur:2004ti, Deur:2008ej, Deur:2014vea, Balla:1997hf}. We present a comparison of various predictions in Fig.~\ref{f2pn}. The results of Refs.\cite{Deur:2004ti, Deur:2008ej, Deur:2014vea} are fitted by using conventional pQCD series for the leading-twist part with fixing $\mu_r=Q$ and the JLab data within different ranges, $0.8~{\rm GeV^2}<Q^2_j<10~{\rm GeV^2}$~\cite{Deur:2004ti}, $0.66~{\rm GeV^2}<Q^2_j<10~{\rm GeV^2}$~\cite{Deur:2008ej} and $0.84~{\rm GeV^2}<Q^2_j<10~{\rm GeV^2}$~\cite{Deur:2014vea}. By using $f^{p-n}_2$, we can evaluate the color polarizability, $\chi_E^{p-n}={2\over3}(2d^{p-n}_2+f^{p-n}_2)$ and $\chi_B^{p-n}={1\over3}(4d^{p-n}_2-f^{p-n}_2)$, which describes the response of the color magnetic and electric fields to the spin of the nucleon~\cite{Ji:1995qe, Stein:1995si}. Using the PMC predictions for the hard-part of the leading-twist contributions, we obtain
\begin{eqnarray}
\chi_B^{p-n}|_{\rm{APT}}&=&0.051\pm0.009, \\
\chi_B^{p-n}|_{\rm{WEB}}&=&0.038\pm0.009, \\
\chi_B^{p-n}|_{\rm{MPT}}&=&0.053\pm0.009, \\
\chi_B^{p-n}|_{\rm{CON}}&=&0.057\pm0.009, \\
\chi_E^{p-n}|_{\rm{APT}}&=&-0.069\pm0.013,\\
\chi_E^{p-n}|_{\rm{WEB}}&=&-0.043\pm0.013,\\
\chi_E^{p-n}|_{\rm{MPT}}&=&-0.075\pm0.013,\\
\chi_E^{p-n}|_{\rm{CON}}&=&-0.082\pm0.013,
\end{eqnarray}
where the errors are squared average of those from $\Delta d^{p-n}_2=\pm0.0036$ and $\Delta f^{p-n}_2$ for the four low-energy $\alpha_s$ models (e.g. Table.~\ref{fitf2mu6}).

\begin{figure*}[htb]
\centering
\subfigure[APT model]{
\includegraphics[width=0.45\textwidth]{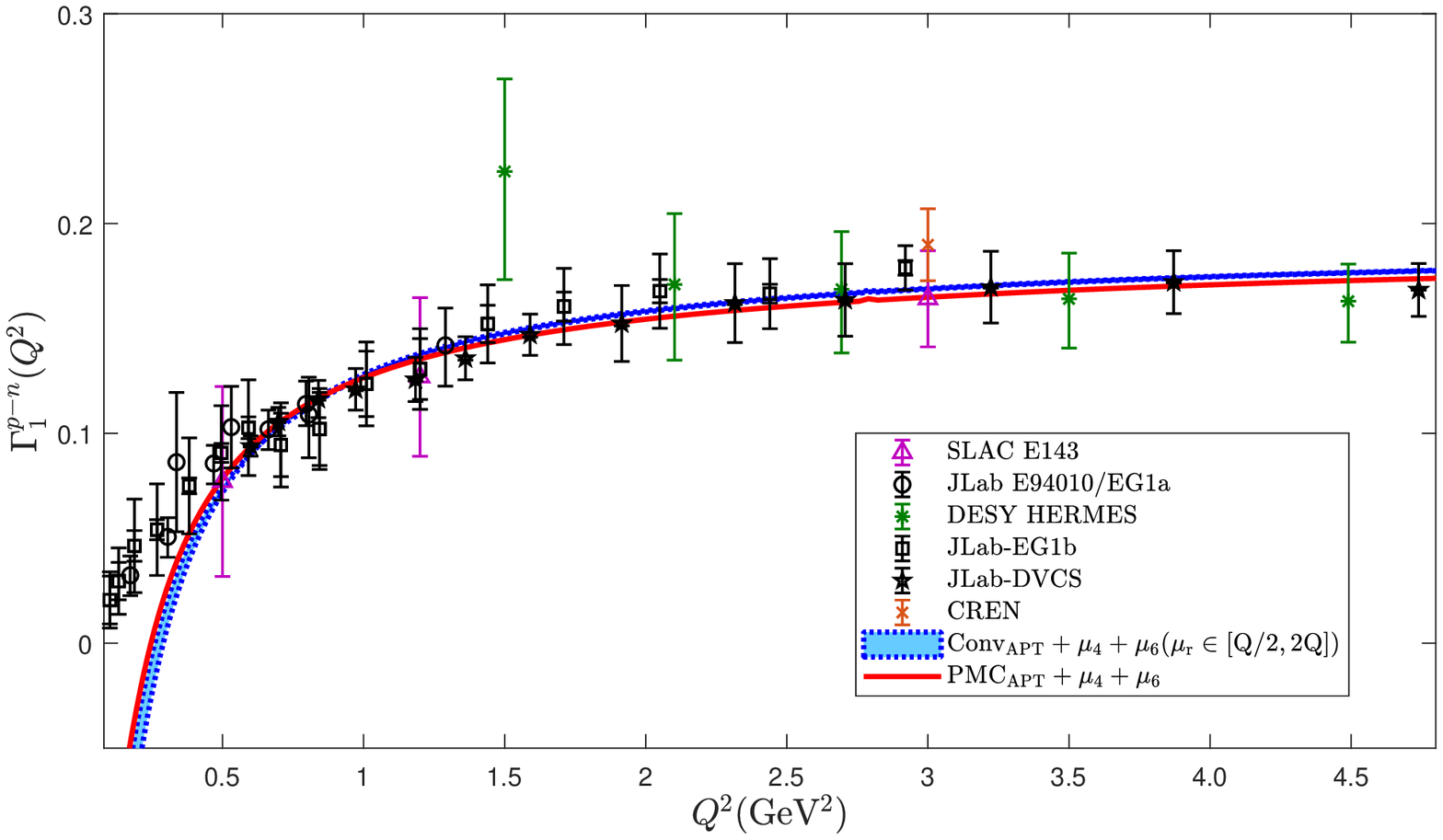}
}
\quad
\subfigure[WEB model]{
\includegraphics[width=0.45\textwidth]{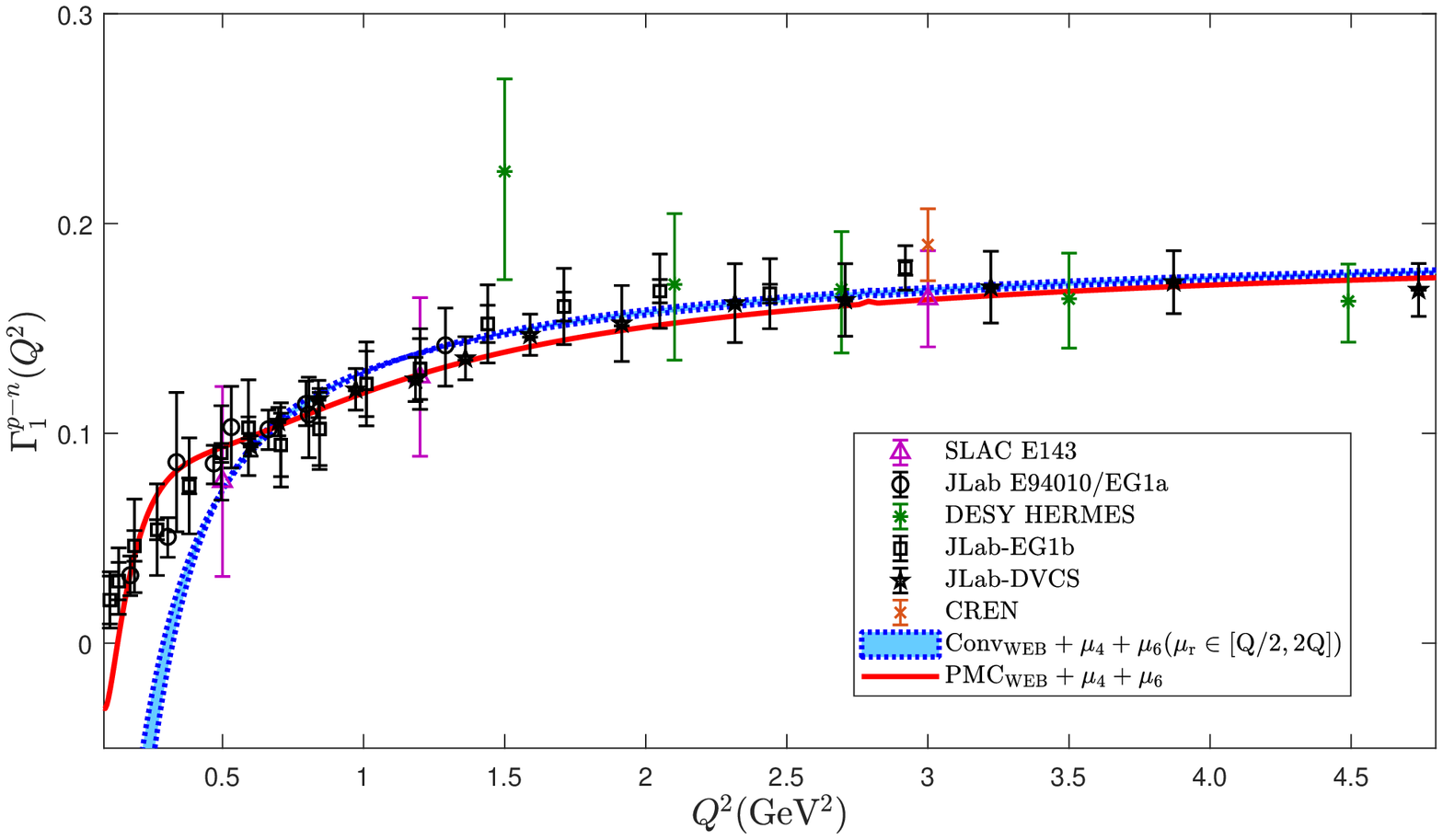}
}
\quad
\subfigure[MPT model]{
\includegraphics[width=0.45\textwidth]{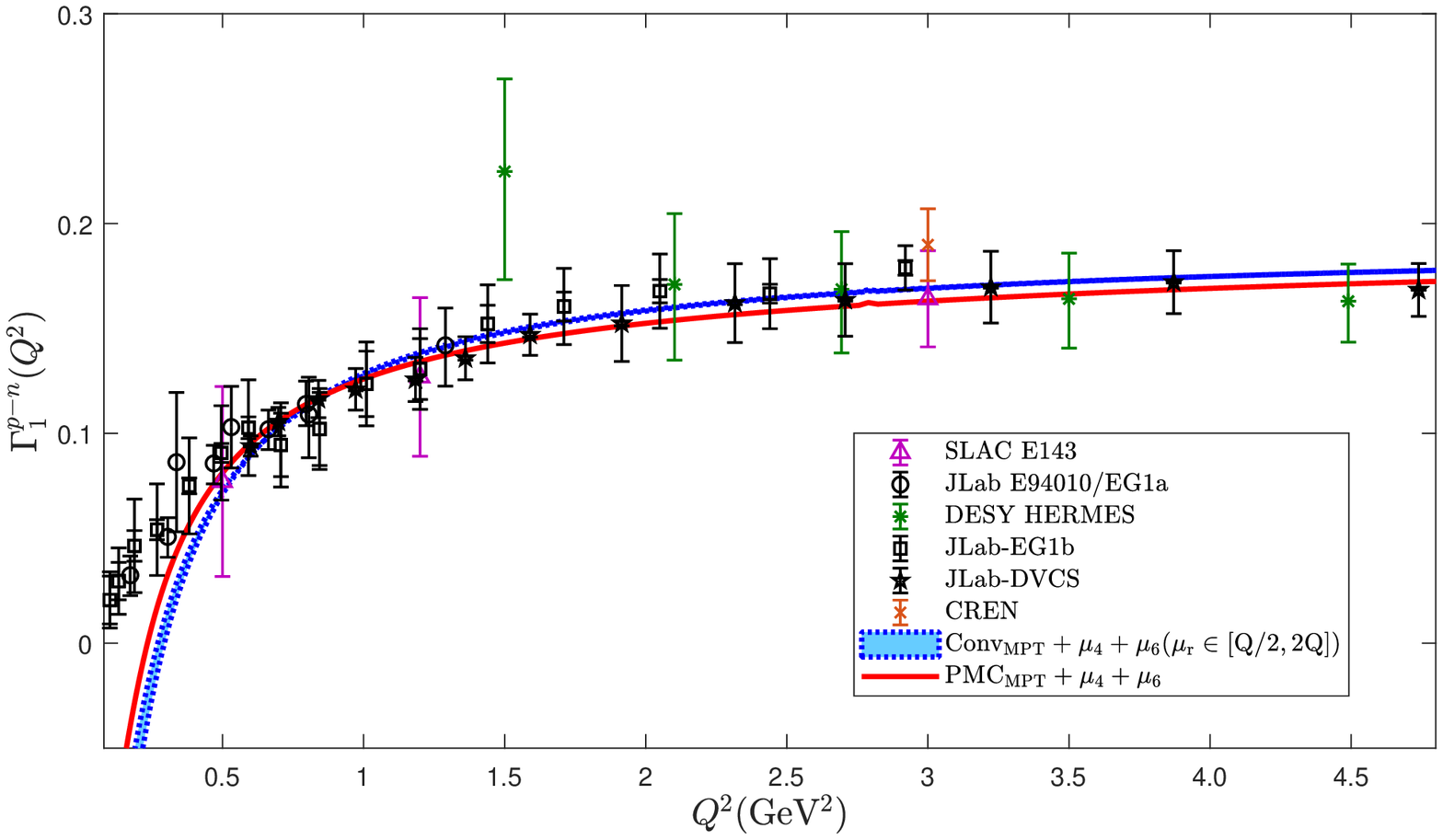}
}
\quad
\subfigure[CON model]{
\includegraphics[width=0.45\textwidth]{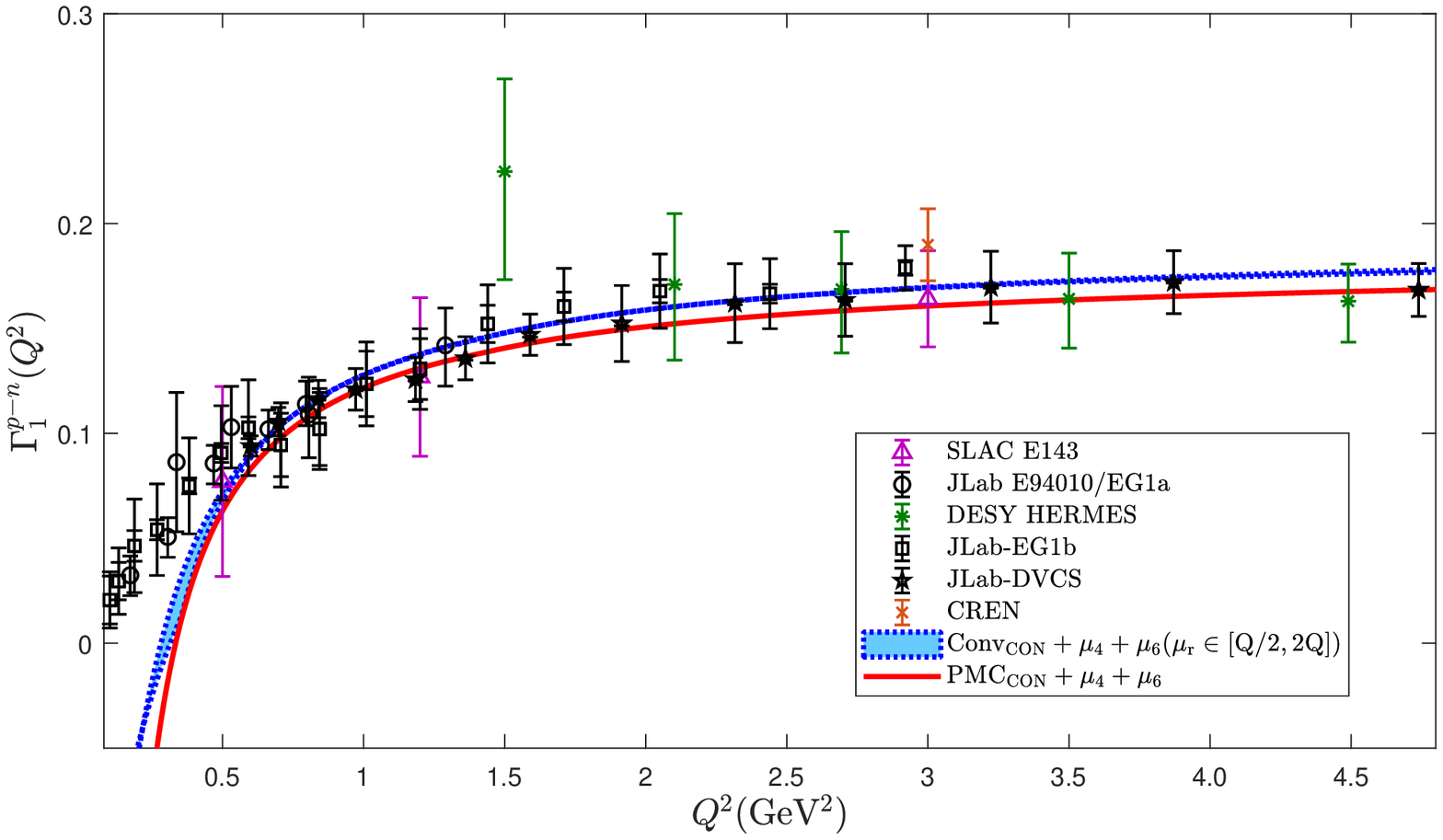}
}
\caption{The spin structure function $\Gamma^{p-n}_1(Q^2)$ with both leading-twist and high-twist contributions under four $\alpha_s$ models: (a) the APT model; (b) the WEB model; (c) the MPT model; and (d) the CON model. The leading-twist perturbative contributions have been calculated up to ${\rm N^3LO}$ level and ${\rm N^4LO}$ level before and after applying the PMC scale-setting approach, respectively. The shaded band shows the prediction under conventional scale-setting approach by varying $\mu_r\in[Q/2,2Q]$. The solid line is the scale-invariant PMC prediction.}
\label{higherBSR}
\end{figure*}

We present the prediction of $\Gamma^{p-n}_1(Q^2)$ with both leading-twist and high-twist contributions in Fig.~\ref{higherBSR}. Comparing with Fig.~\ref{leadingBSR}, Fig.~\ref{higherBSR} shows that a more reasonable prediction can be achieved by including high-twist contributions. Under conventional scale-setting approach, the large scale dependence for the leading-twist prediction of $\Gamma^{p-n}_{1, {\rm Conv.}}(Q^2)$ can be greatly suppressed by including high-twist terms due to the cancellation of scale dependence among different twist-terms. Under PMC scale-setting approach, the scale-invariant $\Gamma^{p-n}_{1, {\rm PMC}}(Q^2)$ under APT, MPT and CON $\alpha_s$ models are close in shape, which as shown by Table.~\ref{fitf2mu6} also have close quality of fit $\chi^2/d.o.f$; while the PMC prediction under WEB model is slightly different from those of other $\alpha_s$ models.

As a final remark, to improve the quality of fit, as suggested by Ref.\cite{Ayala:2018ulm}, we use the JLab data points with $Q^2 > 0.268~{\rm GeV^2}$ to do fit. By using the scale-invariant PMC pQCD series, the quality of fit $\chi^2/d.o.f$ improves to be $\sim 34$ for APT model, $\sim 52$ for WEB model, $\sim 34$ for MPT model and $\sim 38$ for CON model, respectively, which correspond to the $p$-value around $95\%-99\%$~\cite{PDG:2020}.

\subsection{An analysis of high-twist contributions with massive high-twist expression}

As shown by Fig.~\ref{higherBSR}, the predictions drops down quickly in very small $Q^2$-region, and the quality of fit is greatly affected by the data within this $Q^2$-region, indicating the twist-expansion could be failed in very small $Q^2$-region. It has been suggested that by using the ``massive" high-twist expansion to do the data fitting, cf.~\cite{Ayala:2018ulm, Ayala:2020scz, Teryaev:2013qba, Khandramai:2016kbh, Gabdrakhmanov:2017dvg, Aguilar:2014tka}, one may obtain a better explanation of the data in very low $Q^2$ region. As an attempt, we take the following ``massive" high-twist expansion to do the fit~\cite{Ayala:2018ulm}
\begin{eqnarray}
\Gamma^{p-n}_1(Q^2)=\frac{g_A}{6}\left[1-E_{\rm ns}(Q^2)\right]+\frac{\mu^{p-n}_4}{Q^2+m^2}+\cdots,
\label{massivehightwist}
\end{eqnarray}
where the parameter $m$ represents a dynamical effective gluon mass, whose square satisfies
\begin{eqnarray}
m^2=\frac{m^2(1~\rm GeV^2)(1+1/\mathcal{M}^2)^{1+\rm p}}{(1+Q^2/\mathcal{M}^2)^{1+{\rm p}}}.
\end{eqnarray}
Here we have set the initial scale of the squared mass as $1$ GeV, and we shall take the parameters $\mathcal{M}^2=0.5~{\rm GeV}^2$ and $\rm p=0.1$ to do the calculation, which are within the suggested range of Ref.\cite{Aguilar:2014tka}. At present, to fit the magnitude of the ``massive" high-twist terms, the parameters $f^{p-n}_2(1~\rm GeV^2)$ and $m^2(1~\rm GeV^2)$ are used to fit with the data \cite{Deur:2008ej, Deur:2014vea}. When doing the fitting with the experiments data within the range of $0.054\rm GeV^2\le Q^2\le4.739\rm GeV^2$, we adopt four $\alpha_s$ models. The results for the two parameters $f^{p-n}_2(1~\rm GeV^2)$, $m^2(1~\rm GeV^2)$ and their corresponding quality of fit $\chi^2/d.o.f$ are presented in Table.~\ref{fitmassiveBSR}. Those two parameters are obtained by considering the systematic error $\sigma_{j,sys}$ at each data point $Q^2_j$ into the fitting; We put the details of fitting in the end of Appendix B. Comparing the smallest $\chi^2/d.o.f$ listed in Table.~\ref{fitf2mu6} and Table.~\ref{fitmassiveBSR}, one may observes that the ``massive" BSR shows a better behavior with smaller quality of fit $\chi^2/d.o.f$. The conventional predictions for twist-4 $f^{p-n}_2(1~\rm GeV^2)$ apparently depends on the choice of $\mu_r$. The quality of fit $\chi^2/d.o.f$ for conventional predictions with $\alpha_s$ under WEB model varies from tens to hundreds, while similar $\chi^2/d.o.f$ for conventional predictions with $\alpha_s$ under APT, MPT and CON models are along with different fit parameters $f^{p-n}_2(1~\rm GeV^2)$ and $m^2(1~\rm GeV^2)$, respectively. If using the PMC scale-independent series and the ``massive" high-twist term, we can obtain the corresponding color polarizability $\chi^{p-n}_E$ and $\chi^{p-n}_B$:
\begin{eqnarray}
\chi_B^{p-n}|_{\rm{APT}}&=&0.057\pm0.009, \\
\chi_B^{p-n}|_{\rm{WEB}}&=&0.038\pm0.009, \\
\chi_B^{p-n}|_{\rm{MPT}}&=&0.057\pm0.009, \\
\chi_B^{p-n}|_{\rm{CON}}&=&0.060\pm0.009, \\
\chi_E^{p-n}|_{\rm{APT}}&=&-0.083\pm0.014,\\
\chi_E^{p-n}|_{\rm{WEB}}&=&-0.043\pm0.014,\\
\chi_E^{p-n}|_{\rm{MPT}}&=&-0.082\pm0.014,\\
\chi_E^{p-n}|_{\rm{CON}}&=&-0.087\pm0.014,
\end{eqnarray}
where the errors are squared average of those from $\Delta d^{p-n}_2=\pm0.0036$ and $\Delta f^{p-n}_2$ for the four low-energy $\alpha_s$ models (e.g. Table.~\ref{fitmassiveBSR}).

\begin{table*}[htb]
\centering
\begin{tabular}{  c c  c  c c c c c c}
\hline
 & ~$\alpha_s$ models ~   &~   & ~~~$f_2^{p-n}(1)$~~~ &~~~${m}^2(1)$~~~ & ~~~$\chi^2/d.o.f$ \\
\hline
&~                     &$\mu_r=Q/2$          &~~$-0.217\pm{0.004}\pm{0.013}$~~ & ~~$0.203\pm{0.016}\pm{0.102}$~~ &48 \\
&$\rm{APT}|_{Conv}$    &$\mu_r=Q$            &$-0.113\pm{0.004}\pm{0.013}$  &$0.285\pm{0.041}\pm{0.249}$ &42 \\
&~                     &$\mu_r=2Q$           &$-0.166\pm{0.005}\pm{0.013}$ &$0.505\pm{0.045}\pm{0.205}$ &43 \\
&$\rm{APT|_{PMC}}$     &$\mu_r\in[Q/2,2Q]$   &$-0.140\pm{0.004}\pm{0.013}$ &$0.162\pm{0.021}\pm{0.011}$  &28 \\
\hline
&~                     &$\mu_r=Q/2$          &$-0.235\pm{0.004}\pm{0.013}$ &$0.184\pm{0.013}\pm{0.006}$ &37 \\
&$\rm{WEB}|_{Conv}$    &$\mu_r=Q$            &$-0.138\pm{0.009}\pm{0.013}$  &$2.233\pm{0.373}\pm{0.044}$ &80  \\
&~                     &$\mu_r=2Q$           &$-0.183\pm{0.006}\pm{0.013}$ &$0.717\pm{0.056}\pm{0.026}$ &145 \\
&$\rm{WEB|_{PMC}}$     &$\mu_r\in[Q/2,2Q]$   &$-0.081\pm{0.003}\pm{0.013}$ &$0.038\pm{0.010}\pm{0.011}$  & 45 \\
\hline
&~                     &$\mu_r=Q/2$          &$-0.220\pm{0.004}\pm{0.013}$ &$0.220\pm{0.016}\pm{0.046}$ &44 \\
&$\rm{MPT}|_{Conv}$    &$\mu_r=Q$            &$-0.099\pm{0.004}\pm{0.013}$ &$0.229\pm{0.043}\pm{0.044}$ &38\\
&~                     &$\mu_r=2Q$           &$-0.168\pm{0.005}\pm{0.013}$ &$0.538\pm{0.047}\pm{0.058}$ &40\\
&$\rm{MPT|_{PMC}}$     &$\mu_r\in[Q/2,2Q]$   &$-0.139\pm{0.004}\pm{0.013}$ &$0.096\pm{0.015}\pm{0.009}$ &29 \\
\hline
&~                     &$\mu_r=Q/2$         &$-0.215\pm{0.004}\pm{0.013}$ &$0.191\pm{0.016}\pm{0.007}$ &39 \\
&$\rm{CON}|_{Conv}$    &$\mu_r=Q$           &$-0.071\pm{0.003}\pm{0.013}$ &$0.045\pm{0.020}\pm{0.016}$ &60 \\
&~                     &$\mu_r=2Q$          &$-0.174\pm{0.005}\pm{0.013}$ &$0.605\pm{0.052}\pm{0.019}$ &37 \\
&$\rm{CON|_{PMC}}$     &$\mu_r\in[Q/2,2Q]$  &$-0.147\pm{0.004}\pm{0.013}$ &$0.075\pm{0.011}\pm{0.008}$ &36 \\
\hline
\end{tabular}
\caption{The fitted parameters $f^{p-n}_2(Q^2=1~\rm GeV^2)$ and ${m}^2(Q^2=1~\rm GeV^2)$ and their corresponding quality of fit $\chi^2/d.o.f$ under four $\alpha_s$ models before and after applying the PMC, where the first and the second errors are caused by the statistical and systematic errors of the experiments data.}
\label{fitmassiveBSR}
\end{table*}

\begin{figure*}[htb]
\centering
\subfigure[APT model]{\includegraphics[width=0.45\textwidth]{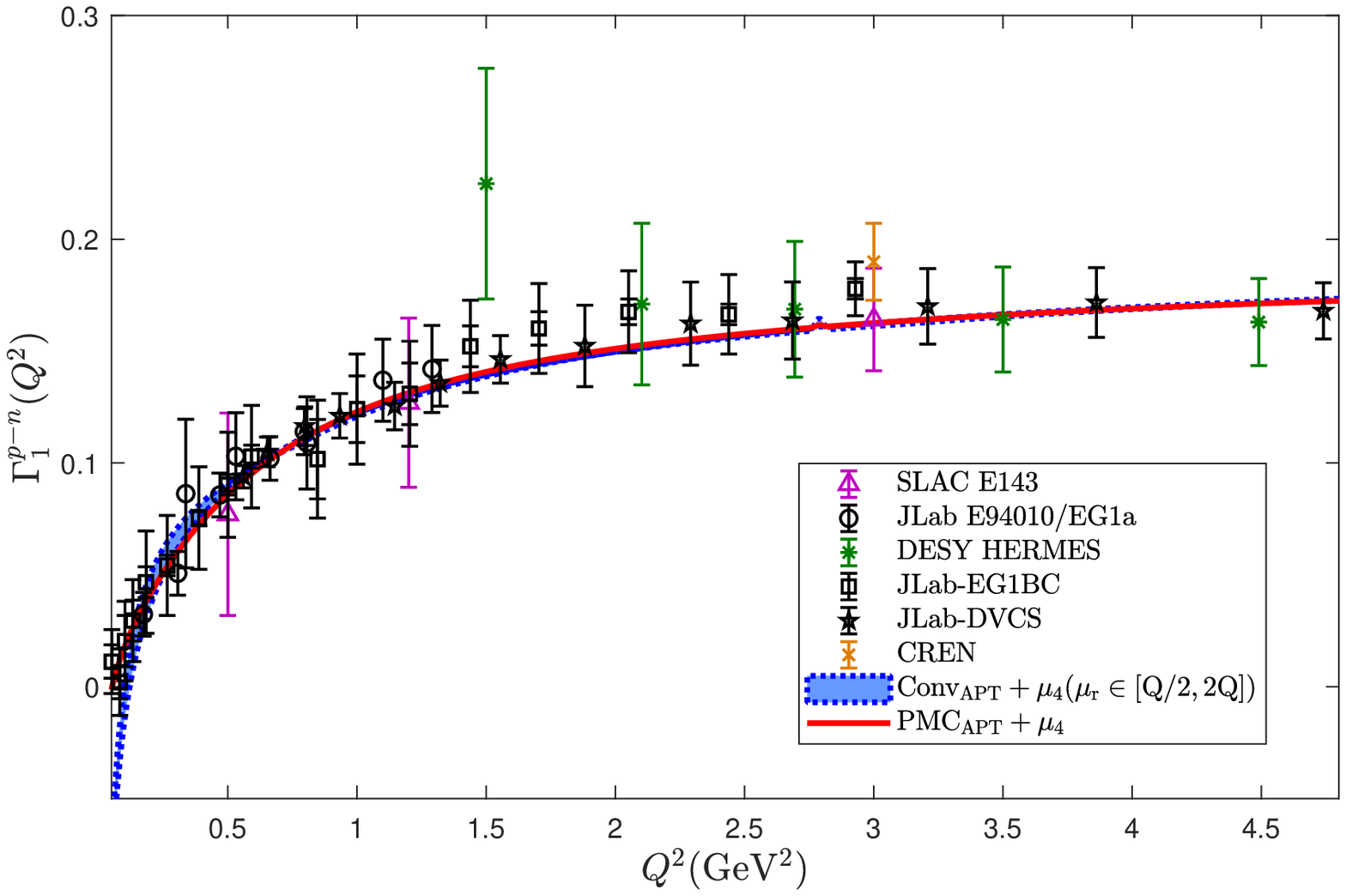}}
\quad
\subfigure[WEB model]{\includegraphics[width=0.45\textwidth]{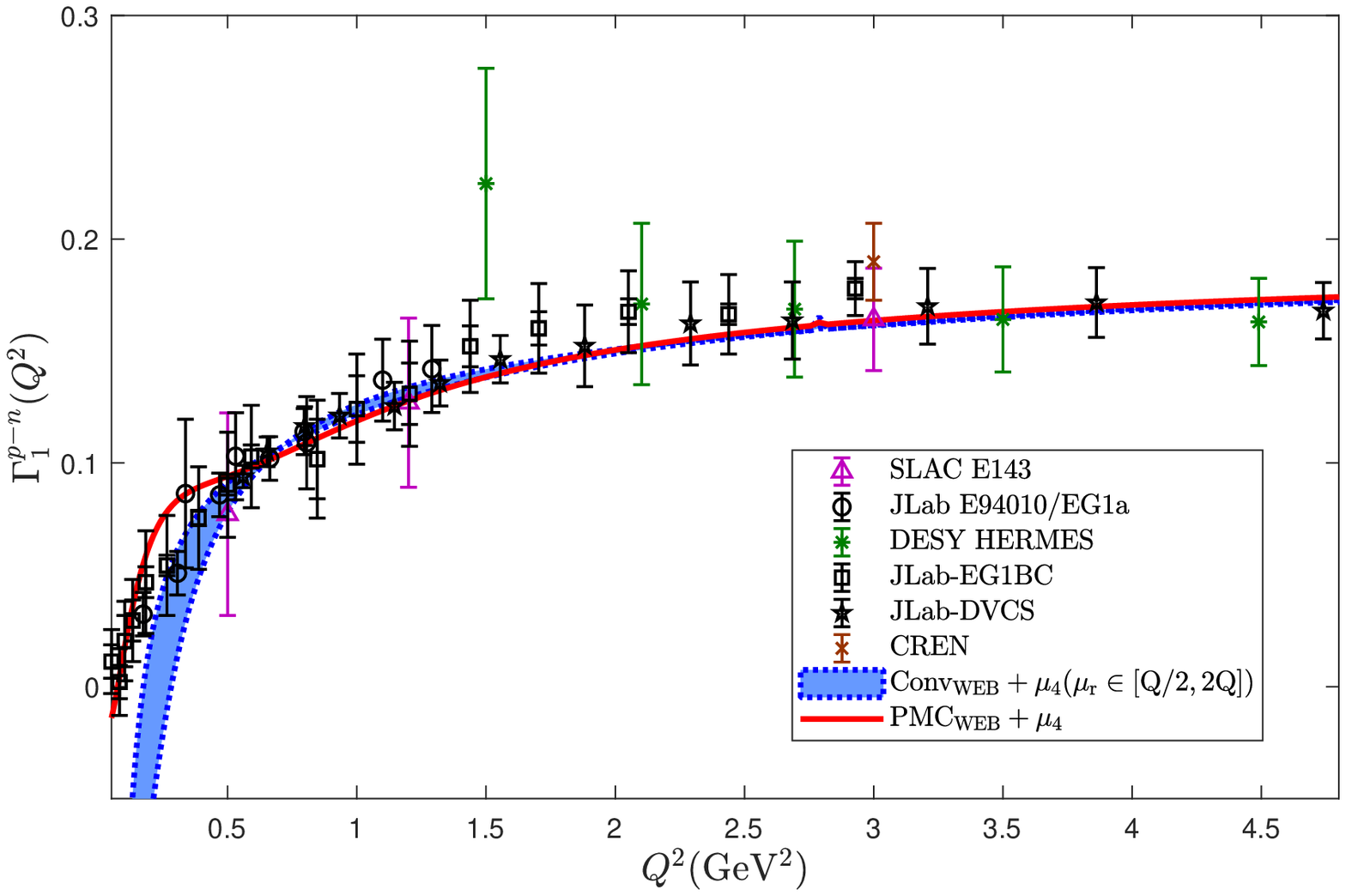}}
\quad
\subfigure[MPT model]{\includegraphics[width=0.45\textwidth]{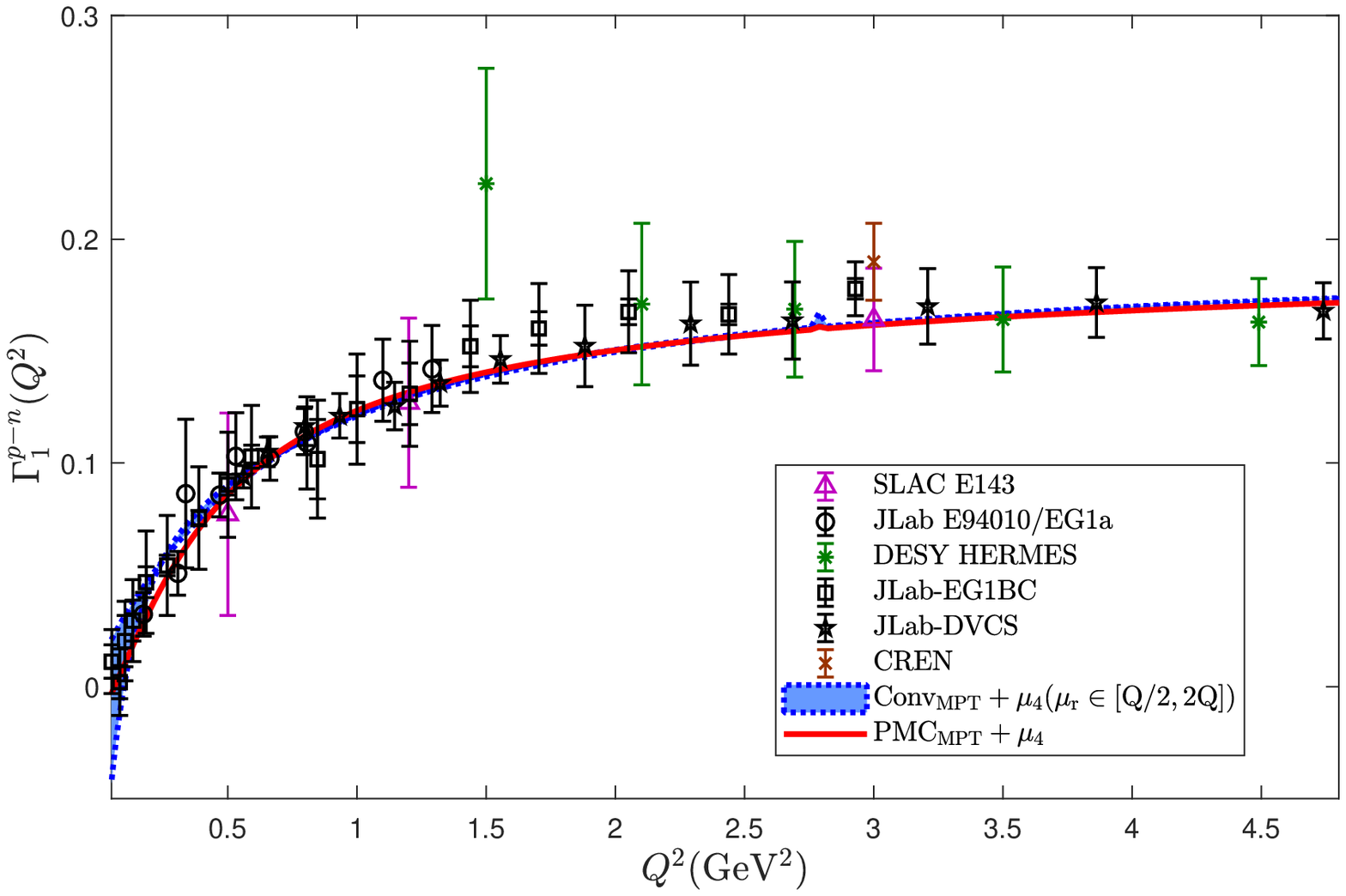}}
\quad
\subfigure[CON model]{\includegraphics[width=0.45\textwidth]{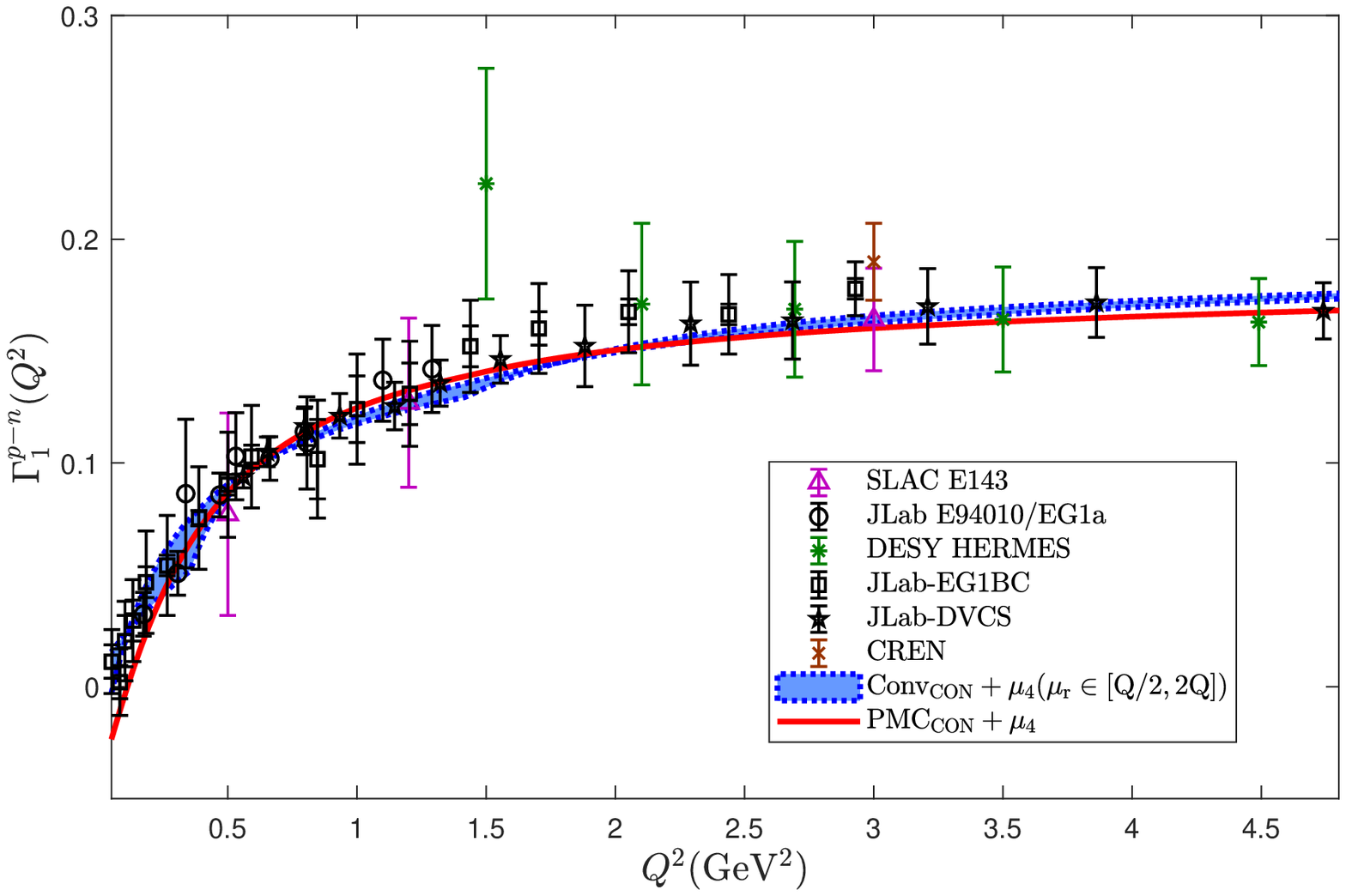}}
\caption{The spin structure function $\Gamma^{p-n}_1(Q^2)$ with both leading-twist and the ``massive" high-twist contributions under four $\alpha_s$ models: (a) the APT model; (b) the WEB model; (c) the MPT model; and (d) the CON model. The leading-twist perturbative contributions have been calculated up to ${\rm N^3LO}$ level and ${\rm N^4LO}$ level before and after applying the PMC scale-setting approach, respectively. The shaded band shows the prediction under conventional scale-setting approach by varying $\mu_r\in[Q/2,2Q]$. The solid line is the scale-invariant PMC prediction.}
\label{massiveBSR}
\end{figure*}

To compare with Fig.~\ref{higherBSR}, Fig.~\ref{massiveBSR} shows that by using the ``massive" high-twist term with the fitted parameters $f^{p-n}_2(1~\rm GeV^2)$ and $m^2(1~\rm GeV^2)$, a better prediction in agreement with the experiments data for $Q^2$ below 0.5 $\rm GeV^2$ can be achieved, which results as a smaller $\chi^2/d.o.f$ in Table.~\ref{fitmassiveBSR}. Different from the PMC predictions, the scale-dependence for conventional predictions is enhanced in small $Q^2$ region. Then, without renormalization scale dependence, the PMC predictions for the twist-4 contribution are more reliable; more explicitly, we observe that the quality of fit $\chi^2/d.o.f$ can be improved as $\sim 28$ for APT model, $\sim 45$ for WEB model, $\sim 29$ for MPT model and $\sim 36$ for CON model, respectively, all of which correspond to a $p$-value $\ge99\%$.

\section{Summary}

In the paper, we have applied the PMC single-scale approach to deal with the perturbative series of the leading-twist part of $\Gamma^{p-n}_{1}(Q^2)$ up to ${\rm N^3LO}$ level. The pQCD series for both $\Gamma^{p-n}_{1}(Q^2)$ and the PMC scale $Q_*$ are convergent in large $Q^2$-region. We have also provided a prediction on the uncalculated ${\rm N^4LO}$ by using the more convergent and scheme-and-scale invariant PMC conformal series. Thus a more accurate pQCD prediction on $\Gamma^{p-n}_{1}(Q^2)$ can be achieved by applying the PMC.

Basing on the PMC predictions on the perturbative part, we then provide a novel determination of the high-twist contributions by using the JLab data, whose momentum transfer lies in the range of ${0.054}~{\rm GeV^2} \leq Q^2\leq 4.739~{\rm GeV^2}$. In large $Q^2$-region, the high-twist contributions to $\Gamma^{p-n}_{1}(Q^2)$ are power suppressed and negligible, which are however sizable in low and intermediate $Q^2$-region; Fig.~\ref{leadingBSR} shows that in low $Q^2$-region, the leading-twist terms alone cannot explain the JLab data. The high-twist term is necessary and it can fix this problem with two fit parameters as Fig.~\ref{massiveBSR} shows. Taking the high-twist contributions up to twist-6 accuracy, we have fixed the twist-4 coefficient $f_2^{p-n}$ and the twist-6 coefficient $\mu_6$ by using four typical $\alpha_s$-models, which give $f_2^{p-n}|_{\rm APT}=-0.120\pm0.013$ and $\mu_6|_{\rm APT}=0.003\pm{0.000}$, $f_2^{p-n}|_{\rm WEB}=-0.081\pm0.013$ and $\mu_6|_{\rm WEB}=0.001\pm{0.000}$, $f_2^{p-n}|_{\rm MPT}=-0.128\pm0.013$ and $\mu_6|_{\rm MPT}=0.003\pm{0.000}$, $f_2^{p-n}|_{\rm CON}=-0.139\pm0.013$ and $\mu_6|_{\rm CON}=0.002\pm0.000$, respectively. Here the errors are squared averages of those from the statistical and systematic errors of the measured data. As an attempt, by taking the ``massive" high-twist expansion such as Eq.(\ref{massivehightwist}) to do the fit, we have shown that a better explanation of the data in very low $Q^2$ range can be achieved.

\hspace{2cm}

\noindent {\bf Acknowledgments:} This work was supported in part by the Natural Science Foundation of China under Grant No.11625520 and No.12047564, by graduate research and innovation foundation of Chongqing, China (Grant No.CYB21045), by the Fundamental Research Funds for the Central Universities under Grant No.2020CQJQY-Z003, and by the Chongqing Graduate Research and Innovation Foundation under Grant No.ydstd1912.

\appendix

\section{the reduced perturbative coefficients ${\hat r}_{i,j}$}

In this appendix, we give the required reduced coefficients $\hat{r}_{i,j}$ for the perturbative series of the leading-twist part of $\Gamma^{p-n}_{1}(Q^2, \mu_r)$ up to four-loop level, i.e.,
\begin{eqnarray}
{\hat r}_{1,0} &=& \frac{3}{4}{\gamma^{\rm ns}_1}, \nonumber \\
{\hat r}_{2,0} &=& \frac{3}{4}{\gamma^{\rm ns}_2}-\frac{9}{16}\big({\gamma^{\rm ns}_1}\big)^2, \nonumber \\
{\hat r}_{2,1} &=& {3 \over 4}{\Pi^{\rm ns}_1}+{K^{\rm ns}_1},  \nonumber\\
{\hat r}_{3,0} &=& \frac{3}{4}{\gamma^{\rm ns}_3}-\frac{9}{8}{\gamma^{\rm ns}_2}{\gamma^{\rm ns}_1}+\frac{27}{64}\big({\gamma^{\rm ns}_1}\big)^3,\nonumber  \\
{\hat r}_{3,1} &=& {3 \over 4}{\Pi^{\rm ns}_2}+{1 \over 2}{K^{\rm ns}_2}-\frac{\gamma^{\rm ns}_1}{4}\left(\frac{3}{2}{K^{\rm ns}_1}+{9 \over 4}{\Pi^{\rm ns}_1}\right), \nonumber \\
{\hat r}_{3,2} &=&0,  \nonumber\\
{\hat r}_{4,0} &=& \frac{3}{4}{\gamma^{\rm ns}_4}-\frac{9}{8}{\gamma^{\rm ns}_3}{\gamma^{\rm ns}_1}-\frac{9}{16}\big({\gamma^{\rm ns}_2}\big)^2 \nonumber\\
 && +  \frac{81}{64} {\gamma^{\rm ns}_2} \big({\gamma^{\rm ns}_1}\big)^2-\frac{81}{256}\big({\gamma^{\rm ns}_1}\big)^4, \nonumber\\
{\hat r}_{4,1} &=& {3 \over 4}{\Pi^{\rm ns}_3}+{1 \over 3}{K^{\rm ns}_3}-{1 \over 4}{\gamma^{\rm ns}_1}\left({K^{\rm ns}_2}+3{\Pi^{\rm ns}_2}\right) \nonumber\\
&& -\frac{\gamma^{\rm ns}_2}{4}\left({K^{\rm ns}_1}+{3 \over 2}{\Pi^{\rm ns}_1}\right) +\frac{\big({\gamma^{\rm ns}_1}\big)^2}{16} \left(3{K^{\rm ns}_1}+\frac{27}{4}{\Pi^{\rm ns}_1}\right),  \nonumber\\
{\hat r}_{4,2} &=& -\frac{3}{16}\big({\Pi^{\rm ns}_1}\big)^2-{1 \over 4}{K^{\rm ns}_1} {\Pi^{\rm ns}_1}, \nonumber\\
{\hat r}_{4,3} &=& 0, \nonumber
\end{eqnarray}
where $\gamma^{\rm ns}_i$, $\Pi^{\rm ns}_i$ and $K^{\rm ns}_i$ can be found in Refs.\cite{Baikov:2010je, Baikov:2012zm}.

\section{Derivation of the parameters $f^{p-n}_2(1~{\rm GeV^2})$ and $\mu_6$}

According to Ref.\cite{Ayala:2018ulm}, it is straightforward to deduce the squares of the standard deviation of $f^{p-n}_2(1{~\rm GeV^2})$ and $\mu_6$, based on the minimization of $\chi^2/d.o.f$. Comparing the experimental data $\Gamma^{p-n}_{1, {\rm exp}}(Q^2)$ and theoretical prediction $\Gamma^{p-n}_{1, {\rm the}}(Q^2)$, we describe this difference at a specific point $Q^2_j$ by using the following symbol:
\begin{eqnarray}
y_j\equiv\Gamma^{p-n}_{1, {\rm exp}}(Q^2_j)-\Gamma^{p-n}_{1, {\rm the}}(Q^2_j).
\end{eqnarray}

The quality parameter $\chi^2$ is rewritten as
\begin{eqnarray}
\chi^2(\mu_4,\mu_6)=\sum_j w_j(y_j-\mu_4 z_j-\mu_6 z^2_j)^2.
\end{eqnarray}
where $z_j\equiv1/Q^2_j$ and $w_j\equiv1/\sigma^2_{j,{\rm stat}}$ from the squared statistical uncertainties of experimental values. The values $\hat{\mu}_4$ and $\hat{\mu}_6$ can be obtained by the condition: the simultaneous minimization of $\chi^2(\mu_4,\mu_6)$. Here the reduced values of $\hat{\mu}_4$ and $\hat{\mu}_6$ are defined as
\begin{eqnarray}
\hat{\mu}_4=\frac{-\overline{y z^2}~\overline{z^3}+\overline{y z}~\overline{z^4}}{\overline{z^2}~\overline{z^4}-\overline{z^3}~\overline{z^3}},
\hat{\mu}_6=\frac{\overline{y z^2}~\overline{z^2}-\overline{y z}~\overline{z^3}}{\overline{z^2}~\overline{z^4}-\overline{z^3}~\overline{z^3}}
\end{eqnarray}
with the unnormalized ``average"
\begin{eqnarray}
 \overline{A}\equiv\sum_j w_j A(z_j).
\end{eqnarray}

Simplifying $f^{p-n}_2$ and $\mu_6$ as constants, there are the following approximations: the statistical uncertainties at different points are considered as uncorrelated; the systematical uncertainties at different point are considered as uncorrelated between different experiments but correlated under the same experiment.

After using Taylor expansion of $\chi^2(\mu_4,\mu_6)$ around the point $(\hat{\mu}_4,\hat{\mu}_6)$ up to the terms quadratic in the deviations, the approximate relations are~\cite{Ayala:2018ulm}
\begin{eqnarray}
\chi^2(\hat{\mu}_4+\sigma(\hat{\mu}^{\rm stat}_4),\hat{\mu}_6)=\chi^2_{\rm min} +\frac{\overline{z^2}~\overline{z^4}}{\overline{z^2}~\overline{z^4}-\overline{z^3}~\overline{z^3}},
\label{statmu4}
\end{eqnarray}
\begin{eqnarray}
\chi^2(\hat{\mu}_4,\hat{\mu}_6+\sigma(\hat{\mu}^{\rm stat}_6))=\chi^2_{\rm min} +\frac{\overline{z^2}~\overline{z^4}}{\overline{z^2}~\overline{z^4}-\overline{z^3}~\overline{z^3}}.
\label{statmu6}
\end{eqnarray}
Thus, the statistical uncertainties of fits parameters $\mu_4$ and $\mu_6$ can be obtained from Eqs.(\ref{statmu4},\ref{statmu6}), and $\sigma(\hat{f}^{stat}_2)=\frac{9}{4 M^2}\sigma(\hat{\mu}^{stat}_4)$ from Eq.(\ref{twist4}).

Moreover, the calculation of systematical uncertainties of the parameters $\mu_4$ and $\mu_6$ at different point should consider the weighted mean values of different experiments. From this, we firstly express the quantities $\overline{y z^2}$ and $\overline{y z}$ in form of $\hat{\mu}_4$ and $\hat{\mu}_6$
\begin{eqnarray}
\overline{y z}=\hat{\mu}_4\overline{z^2}+\hat{\mu}_6\overline{z^3},
\overline{y z^2}=\hat{\mu}_4\overline{z^3}+\hat{\mu}_6\overline{z^4},
\end{eqnarray}
Considering the two different experimental group from Refs.\cite{Deur:2008ej, Deur:2014vea}, the $\hat{\mu}_4$ and $\hat{\mu}_6$ redefined by
\begin{eqnarray}
\hat{\mu}_4&=&\tilde{\mu}^{(1)}_4+\tilde{\mu}^{(2)}_4,\\
\hat{\mu}_6&=&\tilde{\mu}^{(1)}_6+\tilde{\mu}^{(2)}_6,
\end{eqnarray}
and
\begin{eqnarray}
\tilde{\mu}^{(i)}_4&=&\tilde{\alpha}_i\hat{\mu}^{(i)}_4-\tilde{k}_i\hat{\mu}^{(i)}_6,\\
\tilde{\mu}^{(i)}_6&=&\tilde{\beta}_i\hat{\mu}^{(i)}_6+\tilde{h}_i\hat{\mu}^{(i)}_4.
\end{eqnarray}
where $i=1,2$ and weighted factors $\tilde{\alpha}_i$, $\tilde{\beta}_i$, $\tilde{k}_i$ and $\tilde{h}_i$ satisfying that
\begin{eqnarray}
\tilde{\alpha}_i&=&\frac{1}{D^{\rm all}}(\sum^2_{j=1} D^{(ij)}), \\
\tilde{\beta}_j&=&\frac{1}{D^{\rm all}}(\sum^2_{i=1} D^{(ij)}),\\
\tilde{k}_i&=&\frac{1}{D^{\rm all}}\sum^2_{j=1:j\neq i}(-\overline{z^3}^{(i)}~\overline{z^4}^{(j)}
+\overline{z^3}^{(j)}~\overline{z^4}^{(i)}),\\
\tilde{h}_i&=&\frac{1}{D^{\rm all}}\sum^2_{j=1:j\neq i}(-\overline{z^2}^{(i)}~\overline{z^3}^{(j)}
+\overline{z^2}^{(j)}~\overline{z^3}^{(i)}),\\
D^{(ij)}&=&\overline{z^2}^{(i)}~\overline{z^4}^{(j)}-\overline{z^3}^{(i)}~\overline{z^3}^{(j)} ~~(i,j=1,2),\\
D^{\rm all}&=&\sum^2_{j=1}\sum^2_{i=1}D^{(ij)}=\overline{z^2}~\overline{z^4}-\overline{z^3}\overline{z^3}.
\end{eqnarray}
The $D^{\rm all}$ is the unnormalized averages over two experiments. Then, we estimate the systematical uncertainty by averaging the deviations
\begin{eqnarray}
\Delta\tilde{\mu}^{(i),{\rm sys}}_N &&\equiv \sigma(\tilde{\mu}^{(i),sys}_N)\nonumber\\
&&\approx\frac{1}{2}
(|\tilde{\mu}^{(i)}_N({\rm UP})-\tilde{\mu}^{(i)}_N|+|\tilde{\mu}^{(i)}_N({\rm DO})-\tilde{\mu}^{(i)}_N|),
\end{eqnarray}
where $N=4,6$, symbols ``UP" and ``DO" refer to the values $\tilde{\mu}^{(i)}_N$ extracted from the experimental data plus or minus the uncertainty $\sigma_{j, {\rm sys}}$ at momentum $Q_j$, respectively. Finally, the systematical uncertainty $\Delta\hat{\mu}^{\rm sys}_N$ extracted from independent experiments are
\begin{eqnarray}
\Delta\hat{\mu}^{\rm sys}_4 &\equiv& \sigma(\tilde{\mu}^{\rm sys}_4)= \bigg[\sum^2_{i=1}\sigma^2(\tilde{\mu}^{(i),sys}_4)\bigg]^{1/2},\\
\Delta\hat{\mu}^{\rm sys}_6 &\equiv& \sigma(\tilde{\mu}^{\rm sys}_6)= \bigg[\sum^2_{i=1}\sigma^2(\tilde{\mu}^{(i),sys}_6)\bigg]^{1/2},
\end{eqnarray}
and the corresponding uncertainties for the parameter $f^{p-n}_2(1~\rm GeV^2)$ is
\begin{eqnarray}
\Delta\hat{f}^{\rm sys}_2=\sigma(\hat{f}^{\rm sys}_2)=\frac{9}{4 M^2}\sigma(\hat{\mu}^{\rm sys}_4).
\label{sysf2pn}
\end{eqnarray}

When using the ``massive" high-twist expression (\ref{massivehightwist}), the squared standard deviation of $f^{p-n}_2(1{~\rm GeV^2})$ and $m^2(1{~\rm GeV^2})$ can be derived with the help of the minimization of $\chi^2/d.o.f$. If we expand the ``massive" high-twist term in powers of $1/Q^{2}$, the twist-6 term can be expressed as
\begin{eqnarray}
\mu_6(m^2)=-m^2\mu_4;~m^2=-\frac{\mu_6}{\mu_4}.
\end{eqnarray}

Using the approximate relations (\ref{statmu4}, \ref{statmu6}), the statistical uncertainties of the extracted $f^{p-n}_2(1{~\rm GeV^2})$ and $m^2(1{~\rm GeV^2})$ are obtained from the following relations:
\begin{eqnarray}
\chi^2(\hat{\mu}_4+\sigma(\hat{\mu}^{\rm stat}_4),\hat{m}^2)&=&\chi^2_{\rm min} +\frac{\overline{z^2}~\overline{z^4}}{\overline{z^2}~\overline{z^4}-\overline{z^3}~\overline{z^3}},\\
\chi^2(\hat{\mu}_4,\hat{m}^2+\sigma(\hat{m}^2_{\rm stat}))&=&\chi^2_{\rm min} +\frac{\overline{z^2}~\overline{z^4}}{\overline{z^2}~\overline{z^4}-\overline{z^3}~\overline{z^3}}.
\label{statm2}
\end{eqnarray}

As for the systematic uncertainties of the ``massive" case, i.e. the systematic uncertainty of $f^{p-n}_2(1{~\rm GeV^2})$ can obtained from Eq.(\ref{sysf2pn}) and the systematic uncertainty of $m^2(1{~\rm GeV^2})$ can be approximated by the following equations:
\begin{eqnarray}
\sigma(\hat{m}^2)_{\rm sys}&\sim & \bigg(\frac{\hat{\mu}_6}{\hat{\mu}^2_4}\bigg)^2\sigma^2(\hat{\mu}_4)_{sys}+\frac{1}{\hat{\mu}^2_4}\sigma^2(\hat{\mu}_6)_{sys}\nonumber\\
&-&2\bigg(\frac{\hat{\mu}_6}{\hat{\mu}^3_4}\bigg)<\delta\hat{\mu}_4\delta\hat{\mu}_6>_{sys},\\
\label{sysm2pn}
<\delta\hat{\mu}_4\delta\hat{\mu}_6>_{sys}&=&\frac{1}{2}\sum^2_{i=1}\bigg[(\tilde{\mu}^{i}_4({\rm UP})-\tilde{\mu}^{i}_4)(\tilde{\mu}^{i}_6({\rm UP})-\tilde{\mu}^{i}_6)\nonumber\\
&+&(\tilde{\mu}^{i}_4({\rm DO})-\tilde{\mu}^{i}_4)(\tilde{\mu}^{i}_6({\rm DO})-\tilde{\mu}^{i}_6)\bigg].
\end{eqnarray}

\end{document}